\documentclass[journal,comsoc]{IEEEtran}
\usepackage[T1]{fontenc}
\usepackage{bm}
\usepackage{amsthm}
\usepackage{amsmath}
\usepackage{amsfonts}         
\usepackage[cmintegrals]{newtxmath}
\usepackage{cite}
\usepackage{graphicx}
\usepackage{textcomp}
\usepackage[caption=false,font=footnotesize]{subfig}

\newtheorem{definition}{Definition}[section]
\newtheorem{theorem}{Theorem}[section]
\newtheorem{lemma}{Lemma}[section]
\newtheorem{corollary}{Corollary}[section]

\usepackage{hyperref}
\hypersetup{
	colorlinks=true,
	linkcolor=blue,   
	urlcolor=cyan,
	citecolor=green    
}

\begin{document}
\title{Convergence Analysis of Weighted Median Opinion Dynamics with Higher-Order Effects}
\author{Lingrui~Chen,
		Xu~Zhang,~\IEEEmembership{Student~Member,~IEEE,}
		Fanpeng~Song, 
		Fang~Wang, 
		Cunquan~Qu,~\IEEEmembership{Member,~IEEE,} 
		Zhixin~Liu,~\IEEEmembership{Member,~IEEE,}}
\maketitle

\begin{abstract}
The weighted median mechanism provides a robust alternative to weighted averaging in opinion dynamics. Existing models, however, are predominantly formulated on pairwise interaction graphs, which limits their ability to represent higher-order environmental effects. In this work, a generalized weighted median opinion dynamics model is proposed by incorporating high-order interactions through a simplicial complex representation. The resulting dynamics are formulated as a nonlinear discrete-time system with synchronous opinion updates, in which intrinsic agent interactions and external environmental influences are jointly modeled. Sufficient conditions for asymptotic consensus are established for heterogeneous systems composed of opinionated and unopinionated agents. For homogeneous opinionated systems, convergence and convergence rates are rigorously analyzed using the Banach fixed-point theorem. Theoretical results demonstrate the stability of the proposed dynamics under mild conditions, and numerical simulations are provided to corroborate the analysis. This work extends median-based opinion dynamics to high-order interaction settings and provides a system-level framework for stability and consensus analysis.
\end{abstract}


%
\IEEEpeerreviewmaketitle

\section{Introduction}
\label{sec:introduction}
To advance the modeling and analysis of public opinion formation and evolutionary mechanisms, dissect the intrinsic and extrinsic drivers governing opinion change, and unravel the fundamental principles underlying consensus emergence in complex social systems, researchers have increasingly leveraged mathematical modeling and computational simulation techniques to interrogate opinion dynamics\cite{shen_hybrid_2025,wang_trust_2024,liu_opinion_2023,zha_opinion_2020,baumann_modeling_2020,dong_survey_2018,proskurnikov_opinion_2016,jia_opinion_2015}. By integrating empirical data and observational findings, these analytical frameworks facilitate systematic explanation and quantitative prediction of public opinion’s evolutionary trajectories—thus establishing Opinion Dynamics as a rigorous interdisciplinary field bridging engineering, computer science, social science, and systems theory.

Building upon the foundational French model\cite{french1956formal} and the classical DeGroot model\cite{degroot1974reaching}, a wealth of opinion dynamics models have been successively proposed to address evolving research demands\cite{friedkin1990social,altafini2012consensus,hegselmann2002opinion,deffuant2000mixing,dandekar2013biased,bizyaeva2023nonlinear,abelson1964mathematical,axelrod1997dissemination}. The majority of these models adopt complex networks as the core mathematical framework to characterize agent interactions, wherein individual opinions are updated via weighted averaging of neighboring agents’ opinions\cite{ma2025modeling,semonsen2018opinion,he2020opinion,amelkin2019fighting,xia2020expressed}.

However, the widely adopted weighted-averaging mechanism inherently assumes that a larger opinion distance induces a stronger attractive effect. Mei et al. proposed a weighted-median opinion dynamics model, introducing a novel microscopic opinion updating paradigm for opinion dynamics\cite{mei2022micro}. Compared with conventional weighted-averaging mechanisms, this approach more effectively explains opinion diversity in real-world social systems. Mei et al. further established the opinion convergence property of the weighted-median mechanism under asynchronous updating\cite{mei2024convergence}. Complementarily, Zhang et al. proved its convergence characteristics for discrete-time synchronous dynamics, addressing both fully and partially prejudiced agent populations\cite{zhang2025convergence}.

When modeling opinion evolution in networked social systems, existing studies predominantly assume that the external drivers of an agent’s opinion change solely originate from pairwise neighbor interactions. These interactions encompass the ``simple effect'' (influence from a single neighbor) and the ``complex effect'' (successive influence from multiple neighbors)\cite{bassett2012collective,melnik2013multi,zhuang2019multistage}, both of which are inherently direct agent-to-agent interactions. However, the pervasive, subtle yet profound influence of the surrounding environment on individual opinion formation—a core focus of opinion dynamics research—remains underaddressed in conventional frameworks. For instance, individuals initially dismissing trendy products may gradually shift to positive consumption attitudes after immersion in peer circles with frequent product praise and demonstrations; those adhering to strict early routines may adopt flexible schedules when adapting to work/social environments where late nights or weekend sleep-ins are normative; and individuals with low environmental awareness often develop pro-sustainable behaviors (e.g., waste sorting, reusable bags, green transportation) under the influence of eco-conscious communities. These observations demonstrate that individual opinions are dynamically shaped by environmental behavioral norms, information flows, and collective attitudes—a ``subtle and imperceptible influence''\cite{christy2017theorizing,bal2011influence,dillard2015enhancing,kawakami2012implicit} that exposes a critical gap in existing models: conventional pairwise direct interactions cannot fully capture the external drivers of opinion change. Thus, integrating ``indirect interactions'' induced by environmental factors is equally imperative.

While complex interactions and environmental interactions both involve multiple agents—often leading to misclassification as identical—they differ fundamentally in essence. Complex interactions are pairwise agent-to-agent interactions, which can be characterized via network node connections. In contrast, environmental interactions denote agent-group interactions, where groups can be represented by higher-order structures such as simplices\cite{aleksandrov1998combinatorial,carlsson2009topology,salnikov2018simplicial,sizemore2019importance,lambiotte2019networks}. Simplicial complexes have demonstrated substantial value in describing the structure\cite{petri2013topological,sizemore2017classification,kartun2019beyond}, functionality\cite{petri2014homological,lord2016insights,lee2012persistent}, and dynamics of complex networks—including structural brain networks\cite{sizemore2018cliques}, protein interaction networks\cite{estrada2018centralities}, semantic networks\cite{sizemore2018knowledge}, and disease propagation networks\cite{iacopini2019simplicial}.

Thus, this work incorporates environmental factors into opinion dynamics modeling, formalizes social groups as simplices, and adopts simplicial complexes as the underlying structure of networked social systems. Notably, when accounting for environmental influences on agents, this work allows for a key phenomenon: agents may still be influenced by specific social groups without participating in the formation of those groups’ environmental opinions. For instance, in practical scenarios, individuals are often influenced by certain groups or organizations—whose environmental opinions frequently drive changes in personal viewpoints—even though they do not contribute to the formation of such group environmental opinions.

To summarize, this study employs simplicial complexes as the underlying structure for modeling, incorporates higher-order effects into the opinion dynamics analytical framework as ``environmental factors'', and leverages the strengths of the Friedkin-Johnsen model\cite{friedkin1990social} and weighted-median mechanisms to develop a discrete-time synchronous-update opinion dynamics model. This model not only enables agents to retain their intrinsic preferences—incorporating ``agent subjectivity'' as an internal driver—but also accounts for two types of external influencing factors: ``direct neighbor interactions'' and ``indirect environmental interactions'', thereby achieving a more comprehensive reproduction of the opinion update process. The main contributions of this work are summarized as follows: 1) An opinion dynamics model is established on simplicial complexes, incorporating ``indirect environmental-agent interaction'' as an external driver; 2) The convergence of a discrete-time synchronous opinion dynamics model adopting the weighted-median mechanism is analyzed under higher-order network structures; 3) Specifically, for the scenario where all agents are opinionated, the system’s convergence and exponential convergence rate are derived; for the scenario where agents are a mix of opinionated and unopinionated, a sufficient condition for the system to achieve asymptotic consensus is provided.

This work proceeds as follows: Section \ref{B2} defines the notation; Section \ref{B3.1} presents the model setup; Section \ref{B3.2} formulates the opinion updating rule; Section \ref{B3.3} focuses on convergence analysis for the model with partially opinionated agents; Section \ref{B3.4} addresses convergence analysis for fully opinionated agents; Section \ref{B4} presents simulation results and analysis; and Section \ref{B5} offers concluding remarks and outlines future research directions. For brevity, proofs omitted from the main text are provided in the Appendix.

\section{Notation}
\label{B2}
1. Notation for Simplicial Complexes: Let $G$ denote a social network, where the connections between nodes represent pairwise interaction relationships. Based on network $G$, a simplicial complex $K_G$ is constructed. Within this framework, $V(K_G)$ stands for the vertex set of the simplicial complex $K_G$, and these vertices correspond to the set of agents in the group; Simp$(K_G)$ denotes the set of all simplices of various dimensions in the simplicial complex $K_G$, which is equivalent to the set of environments in the group under the research context of this work.

2. Mathematical Notation: The notation $\bm{x} \in \mathbb{R}^n$ denotes that $\bm{x}$ belongs to the $n$-dimensional real Euclidean space. The notation $\|\cdot\|_{\infty}$ denotes the standard infinity norm (also known as the Chebyshev norm); for a vector $\bm{x} \in \mathbb{R}^n$, defined as the maximum absolute value of its components. Additionally, $I_n$ denotes the $n \times n$ identity matrix, with ones on the diagonal and zeros elsewhere.

3. Weighted Median: For $\boldsymbol{x} \in \mathbb{R}^n$, let $\boldsymbol{w} \in \mathbb{R}^n$ be the associated weight vector, where $w_i$ ($i = 1, \dots, n$) weights the $i$-th component $x_i$ of $\boldsymbol{x}$. The weighted median of $\boldsymbol{x}$ with respect to $\boldsymbol{w}$, denoted $Med_i(\boldsymbol{x}; \boldsymbol{w})$, is formally defined as follows:
\begin{definition}
	(Weighted Median\cite{mei2022micro,zhang2025convergence}) Let $\bm{x} = (x_1, \dots, x_n)^T \in \mathbb{R}^n$ be a vector with an associated weight vector $\bm{w} = (w_1, \dots, w_n)^T$, where $w_i \ge 0$ and $\sum_{i=1}^{n} w_i = 1$.
	If a value $x^*$ satisfies
	\[
	\sum_{i: x_i < x^*} w_i \le \frac{1}{2}
	\quad \text{and} \quad 
	\sum_{i: x_i > x^*} w_i \le \frac{1}{2},
	\]
	then $x^*$ is called a weighted median of $\bm{x}$ with weights $\bm{w}$.
	\label{D2.1}
\end{definition}

While the weighted median is not necessarily unique, $x^*$ is the unique weighted median of $\bm{x}$ with respect to $\bm{w}$ if it further satisfies:
\begin{equation*}
	\sum_{i:x_{i}<x^{*}}w_{i} < \frac{1}{2}\, , \sum_{i:x_{i}=x^{*}}w_{i} = \frac{1}{2}\,\,\, \text{and} \sum_{i:x_{i}>x^{*}}w_{i} < \frac{1}{2}.
\end{equation*}

\section{Environmental-Impacted Weighted Median Opinion Dynamics}
\label{B3}
This section proceeds as follows: we first formulate the environmental-impacted weighted median opinion dynamics model, and then formalize the corresponding opinion update rule. Subsequently, we analyze the system dynamics with partially opinionated agents, before investigating the scenario with fully opinionated agents.

\subsection{Model Setup}
\label{B3.1}
In practice, agents' opinions evolve gradually under sustained, indirect social context influences—a phenomenon termed the ``subtle and imperceptible influence'' effect. To characterize the general mutual influence among agents, we first model their social interactions via a network, formally defined as $G = (V, E)$ with $V = \{1, 2, \dots, n\}$ denoting the agent set and $E \subseteq V \times V$ the edge set encoding pairwise interaction links between agents. For each agent $i \in V$, the opinion at time $t$ is denoted by $x_i(t) \in \mathbb{R}$. Correspondingly, the system-level opinion vector at time $t$ is given by
\[\bm{x}(t) = \left(x_1(t), x_2(t), \dots, x_n(t)\right)^{\top}.\]

However, relying solely on first-order neighbor interactions fails to fully capture the environmental effects experienced by agents. To address this limitation, we introduce higher-order structures: by abstracting the environment of agent as a simplex, we construct a simplicial complex $K_G$ from the underlying network $G$. The vertex set of $K_G$ coincides with the node set of $G$, i.e., $V(K_G) = V$. Since $V(K_G) \equiv V$, we denote $V(K_G)$ simply as $V$ in subsequent discussions. Let $\text{Simp}(K_G) = \{\delta_1, \delta_2, \dots, \delta_l\}$ denote the set of all simplices in $K_G$, which serves as the environment set. Each environment $\delta_k \subseteq V$ is a subset of $V$, representing a group of agents interconnected within a specific context—e.g., an organizational department, an interest group, or participants in a shared event. For each environment $\delta_k \in \text{Simp}(K_G)$, we denote its opinion at time $t$ as $y_k(t) \in \mathbb{R}$. Correspondingly, the system-wide environmental opinion vector at time $t$ is given by
\[\bm{y}(t) = \left(y_1(t), y_2(t), \dots, y_l(t)\right)^{\top}.\]

In this work, we assume that the environmental opinion vector $\bm{y}(t)$ is formulated as a function of the agent opinion vector $\bm{x}(t)$. Each component of $\bm{y}(t)$ corresponds to the environmental opinion associated with a distinct simplex. Specifically, the environmental opinion of simplex $\delta_k$ is defined as the weighted sum of opinions of all agents residing within this simplex $\delta_k$. 
To formalize this relationship, we first introduce the construction of the indicator matrix for the simplicial complex, whose mathematical expression is given by $\bm{A} = (a_{ki})_{l \times n}$.
The element $a_{ki}$ of matrix $\bm{A}$ is defined as the contribution weight of agent $i$ to simplex $\delta_k$, where a simplex serves as an environmental unit. If $a_{ki} = 0$, this implies that agent $i$ does not belong to simplex $\delta_k$, i.e., $i$ is not a member of $\delta_k$. Directly following the above definition, the sum of elements in each row of $\bm{A}$ is unity, rendering $\bm{A}$ a row-stochastic matrix.
Building on the definition of indicator matrix $\bm{A}$, we formulate the explicit expression for the environmental opinion using its contribution weights.
\begin{equation}
	\bm{y}(t)=\bm{A}\bm{x}(t)
	\label{1.4}.
\end{equation}
\subsection{Opinion Updating Rule}
\label{B3.2}
In this work, the opinions of agents within the proposed framework are dynamically updated in accordance with the following rule:
\begin{equation}    
	x_{i}(t+1) = \lambda_{i}u_{i}+(1-\lambda_{i})E_{i}(\bm{x}(t)),~\forall t \in \mathbb{N}
	\label{1.1},
\end{equation}
where $\lambda_i$ denotes the anchoring coefficient of agent $i$, $u_i$ its intrinsic bias, and $E_i(\bm{x}(t))$ its external opinion. The explicit expression of $E_i(\bm{x}(t))$ is given as follows:
\begin{equation}
	E_{i}(\bm{x}(t))\!=\! (1\!-\!\gamma_{i})Med_{i}(\bm{x}(t);\!\bm{W})\!+\!\gamma_{i}Med_{i}(\bm{A}\bm{x}(t);\!\bm{M})
	\label{1.2},
\end{equation}
where $\gamma_i \in [0,1]$ denotes the environmental sensitivity coefficient of agent $i$, quantifying its responsiveness to external environmental influences.
First-order inter-agent network influence weights are captured by the adjacency matrix $\bm{W} \in \mathbb{R}^{n \times n}$ (row-stochastic), with each entry $w_{ij}$ encoding the direct interaction strength imposed on agent $i$ by agent $j$. Notably, asymmetric interactions are permitted, i.e., $w_{ij} \neq w_{ji}$, reflecting real-world scenarios where influence is not necessarily reciprocal. Meanwhile, higher-order environmental influence weights on agents are encapsulated by the matrix $\bm{M} \in \mathbb{R}^{n \times l}$ (also row-stochastic), wherein each entry $m_{ik}$ quantifies the strength of indirect environmental effects exerted on agent $i$ by the simplex $\delta_k$.
$Med_{i}(\bm{x}(t);\bm{W})$ denotes the weighted median of $\bm{x}(t)$ with respect to the weight vector $\bm{w}_i^T = (w_{i1}, \ldots, w_{in})$. Should the weighted median be non-unique, we define $Med_{i}(\bm{x}(t);\bm{W})$ as the median closest to $x_i(t)$.

In this work, we focus on whether the agents' opinions in system (\ref{1.1}) converge over time—i.e., whether they cease to evolve and attain a stable state. Subsequent sections analyze opinion convergence in the system under distinct scenarios and further investigate the conditions for the system to achieve consensus. Prior to proceeding, we first formalize the definition of consensus attainment.

\begin{definition}\cite[Def. 3.1]{zhang2025convergence}
	For $\forall\, \bm{x}(0) \in \mathbb{R}^{n}$, if there exists a constant $x^{*} \in \mathbb{R}$ such that for all $i \in V$, we have $\lim_{t \to \infty} x_{i}(t) = x^{*}$, then we say that the system (\ref{1.1}) asymptotically achieves consensus.
	\label{D2.2}
\end{definition}

\subsection{Analyzing Partially Opinionated Agents}
\label{B3.3}
In real-world social systems, the innate diversity of individual traits and cognitive styles drives marked heterogeneity in agents’ opinion formation, maintenance, and updating. Take local community forum debates on urban greening policies as an example: some participants (e.g., a retired environmental engineer or a long-term resident advocating for children’s playspaces) hold unwavering views grounded in professional expertise or decades of lived experience, whereas young professionals in attendance tend to listen attentively, endorse compelling arguments, and adjust their stances flexibly without rigid commitments. This dichotomy between agents with entrenched versus malleable opinions is no anecdotal phenomenon but a fundamental property of social networks, manifesting across contexts from workplace decision-making (senior managers often hold firm views; new hires remain adaptable) to online public discourse (opinion leaders versus casual followers).
Formally, we classify these two archetypes as opinionated and unopinionated agents, respectively. Opinionated agents display a strong cognitive anchoring effect: their self-formed opinions act as stable reference points during social interaction, and they only partially revise their views even when faced with conflicting perspectives. In contrast, unopinionated agents lack such cognitive persistence—they embrace external information openly, with their initial opinions serving as transient starting points rather than fixed anchors.

To mathematically formalize this heterogeneous social structure, we partition the vertex set $V$ of $K_G$ into two disjoint subsets, denoted as $V_1 := \{1, 2, \ldots, n_1\} \subseteq V$ and $V_2 := \{n_{1}\!+\!1, n_{1}\!+\!2, \ldots, n_1\!+\! n_2\} \subseteq V$. The subset $V_1$ corresponds to the opinionated agent group, whose opinion updating dynamics are governed by an anchoring coefficient $\lambda_i \in (0, 1]$. This coefficient quantifies the degree of reliance on an intrinsic bias value $u_i$, where $u_i$ may correspond to the agent’s initial opinion or an externally formed stance (e.g., a pre-established belief derived from cultural norms, expert consensus, or prior experience). 
Specifically, a $\lambda_i$ approaching 1 denotes an agent with nearly absolute adherence to $u_i$—e.g., a seasoned scientist upholding empirically grounded, well-verified theoretical frameworks. In contrast, a $\lambda_i$ at the lower end of the interval reflects modest yet meaningful persistence toward $u_i$—e.g., a long-term community resident with firm but adaptable local policy preferences, rooted in long-term life experience yet responsive to new collective needs or practical constraints.
In sharp contrast, $V_2$ represents the unopinionated agent group, for whom the anchoring effect is absent in opinion evolution—accordingly, their anchoring coefficient is set to $\lambda_i = 0$—e.g., a college student researching a controversial topic might initially hold a tentative view but readily revises it when engaging with academic literature, expert insights, and peer deliberations, with no inherent bias toward any pre-defined stance (including their initial perspective).

To enhance analytical tractability and isolate the impact of anchoring heterogeneity, we introduce a simplifying assumption in this section: all agents share a common bias term $u_i = u$ for all $i \in V$. Under this premise, the opinion dynamics system (\ref{1.1}) derived earlier can be re-expressed in the following form:

\begin{equation}
	x_{i}(t+1) = 
	\begin{cases}
		\lambda_{i}u+(1-\lambda_{i})E_{i}(\bm{x}(t)), & i \in V_{1}; \\
		E_{i}(\bm{x}(t)), & i \in V_{2}, 
	\end{cases}
	\label{2.1}
\end{equation}
where $\lambda_i \in (0, 1]$ denotes the anchoring coefficient quantifying each agent’s adherence to the common bias $u$. For system (\ref{2.1}), the core objective of this section is to derive conditions under which the system achieves asymptotic consensus.

Through meticulous observation of the evolutionary dynamics of opinions within diverse social groups in real-world contexts, it becomes feasible to extract inherent structural patterns and further abstract them into operational conceptual frameworks. Hereafter, we delineate four such distinct structures.
\begin{definition}
	(\textit{Cohesive Agent Set}) If there exists a non-empty subset $P\subset V$ such that for any $i\in P$, $\sum_{j\in P}w_{ij}\geq\frac{1}{2}$ holds, then $P$ is called a cohesive individual set of $K_G$.
	\label{D3.1}
\end{definition}

This construct models a highly cohesive subgroup where each member engages in robust internal interactions. For instance, consider a team of researchers conducting long-term collaborative research: each researcher frequently exchanges ideas with intra-team peers, with such interactions accounting for over half of their total social engagement—an interaction pattern that fosters mutual trust and enables consistent information dissemination. A defining characteristic of this subgroup is that its influence is inherently confined to its members, which distinguishes it from broader ``group set'' concepts that typically encompass more extensive spheres of influence.

\begin{definition}
	(\textit{Strong Cohesive Group Set}) If there exists a non-empty subset $Q\subset \text{Simp}(K_G)$, i.e., $Q$ is a set composed of simplices, satisfying the following two conditions:
	
	(i) Each simplex in the set $Q$ is composed of agents in cohesive agent set $P$.
	
	(ii) For $\forall\,i\in V$, it satisfies $\sum_{k\in Q}m_{ik}>\frac{1}{2}$. 
	
	\noindent Then $Q$ is called a strong cohesive group set of $K_G$.
	\label{D3.2}
\end{definition}

Building on the cohesive individual set, this concept denotes a collection of simplices (e.g., collaborative subgroups, joint initiatives) rooted in a cohesive individual set $P$. For instance, consider the aforementioned research team (i.e., $P$, a cohesive individual set): it publishes a series of high-impact joint works—with such collective endeavors constituting the simplex set $Q$. These endeavors exert substantial influence: over half of all researchers in the field—formally represented as $ \forall i \in V $—cite their publications, a pattern formally quantified by $ \sum_{k\in Q}m_{ik}>\frac{1}{2} $. A key distinction between this construct and a weak cohesive group set resides in its foundational anchor: $Q$ is explicitly grounded in a pre-existing cohesive individual set $P$.

\begin{definition}
	(\textit{Weak Cohesive Group Set}) If there exists a non-empty subset $Q\subset \text{Simp}(K_G)$ such that for $\forall\, i\in V$, $\sum_{k\in Q}m_{ik}>\frac{1}{2}$, then $Q$ is called a weak cohesive group set of $K_G$.
	\label{D3.3}
\end{definition}

In contrast to the strong cohesive group set, this set denotes a collection of simplices that exerts influence without anchoring in a preexisting cohesive individual set. A paradigmatic example is a viral social media movement: diverse users (e.g., ordinary citizens, micro-influencers, small organizations) devoid of preexisting formal ties generate and disseminate content around a pressing social issue—with each piece of content or collaborative post constituting a simplex within $Q$. Though devoid of a cohesive core, their decentralized, collective messaging resonates with more than half of all platform users ($\forall i \in V$), shaping public opinion—formally, $\sum_{k\in Q}m_{ik}>\frac{1}{2}$. Its defining characteristic is ``influence without cohesion'': the simplex set $Q$ gains momentum via broad-based participation rather than a tight-knit core.

\begin{definition}
	(\textit{Cohesive Influential Cluster}): If a non-empty subset $P \subset V$ is itself a cohesive agent set, and $P$ is associated with a strong cohesive group set $Q \subset \text{Simp}(K_G)$, then $P$ is called a cohesive influential cluster of $K_G$.
	\label{D3.4}
\end{definition}

This concept synthesizes the cohesive individual set and the strong cohesive group set, establishing an integrated construct that unites their defining features. A quintessential illustration is a leading academic research laboratory (i.e., $P$): its members form a cohesive individual set—characterized by intensive internal collaboration and satisfying the condition $\sum_{j\in P}w_{ij}\geq\frac{1}{2}$ for all $i \in P$—whereas their collective outputs (e.g., co-authored publications, open-source analytical tools) constitute the strong cohesive group set $Q$. This set $Q$ exerts dominant influence over the broader research community, formally quantified by $\sum_{k\in Q}m_{ik}>\frac{1}{2}$ for all relevant researchers $i \in V$. Critically, the cohesive influential cluster embodies two mutually reinforcing defining attributes: ``internal cohesion'', instantiated by the tight-knit collaborative structure of $P$, and ``external influence'', mediated by the community-wide impact of $Q$. This dual nature differentiates the cohesive influential cluster from two distinct counterparts: (1) cohesive individual set, which lack external influence despite internal cohesion; and (2) weak cohesive group set, which lack a preexisting, stable cohesive core even when exerting limited influence.

With the definitions of these four special structures, we now turn to investigating the structural configurations that underpin a system’s capacity to achieve asymptotic consensus. Before presenting the main conclusions, we provide some lemmas.

\begin{lemma}
	Consider the system (\ref{2.1}), the weighted median $Med_{i}(\bm{A}\bm{x};\bm{M})$ satisfies the following inequality:
	\begin{equation}
		\underset{j \in V}{\min}\,x_{j}
		\leq Med_{i}(\bm{Ax};\bm{M})
		\leq \underset{j \in V}{\max}\,x_{j}
	\end{equation}
	for $\forall\, i\in V$ and $\forall\, \bm{x}=(x_{1},x_{2},...,x_{n})^{\top}\in \mathbb{R}^{n}$.
	\label{L3.1}
\end{lemma}

Lemma \ref{L3.1} implys that for any agent $i$ in a group, the weighted median of its associated environment opinion is between the maximum and minimum of all agent opinions in that group.

Having the range of the weighted median of environment opinion, a conclusion is given in the reference \cite{zhang2025convergence} for the range of the weighted median of the agent opinion.

\begin{lemma}
	(Lemma 4.1 of \cite{zhang2025convergence})
	Consider a network composed of $n$ agents, with the influence matrix between agents being $\bm{W}=(w_{ij})_{n\times n}$. If there exists an agent $i\in V$ and a set $P\subset V$ satisfying
	\begin{equation}
		\sum_{j\in P}w_{ij}\begin{cases}
			>\frac{1}{2},\quad i\notin P;\\
			\geq \frac{1}{2},\quad i\in P,
		\end{cases}
	\end{equation}
	then for $\forall\, \bm{x}=(x_{1},x_{2},\ldots, x_{n})^{T}\in \mathbb{R}^{n}$, we have
	\begin{equation}
		\underset{j\in P}{\min}\,x_{j}
		\leq Med_{i}(\bm{x};\bm{W})
		\leq \underset{j\in P}{\max}\,x_{j}.
	\end{equation}
	\label{L3.2}
\end{lemma}

In Lemma \ref{L3.2}, we have identified a key phenomenon: when a specific structure exists in the network, the weighted median exhibits a surprising conclusion of boundedness. 
To deeply explore the attributes of higher-order networks, we first need to introduce several core definitions based on the simplicial complex $K_G$, to lay a theoretical foundation for subsequent research.
Below is an important lemma.

\begin{lemma}
	Consider the system (\ref{2.1}). 
	If there exists a cohesive influential cluster $P\subset V$,
	then for $\forall\, i \in P$, $\, \bm{x}=(x_{1},x_{2},...,x_{n})^{T}\in \mathbb{R}^{n}$ satisfies
	\begin{equation}
		\underset{j\in P}{\min}\,x_{j}
		\leq 
		E_{i}(\bm{x})
		\leq \underset{j\in P}{\max}\,x_{j}
		\label{3.3}.
	\end{equation}
	\label{L3.3}
\end{lemma}

The following lemma illustrates that if there exists a special cohesive influential cluster in a simplicial complex $K_G$, the opinion of an agent will be within a specific range.

\begin{lemma}
	Consider the system (\ref{2.1}), if there exists a cohesive influential cluster $P \subset V$ consisting only of unopinionated agent, then for $\forall\, i \in P,~\, t \in \mathbb{N}$, we have
	\begin{equation}
		\underset{j \in P}{\min}\,x_{j}(0)
		\leq x_{i}(t)
		\leq \underset{j \in P}{\max}\,x_{j}(0).
	\end{equation}
	\label{L3.4}
\end{lemma}

The following two lemmas give the monotonicity in the evolution of opinion.

\begin{lemma}
	For the system (\ref{2.1})\\
	(i) If there exists $T \geq 0$ such that for any $t \geq T$, we have $u \geq \min_{i \in V}x_{i}(t)$, then for $t \geq T$, $\min_{i \in V}x_{i}(t)$ is monotonically non-decreasing.\\
	(ii) If there exists $T \geq 0$ such that for any $t \geq T$, we have $u \leq \max_{i \in V}x_{i}(t)$, then for $t \geq T$, $\max_{i \in V}x_{i}(t)$ is monotonically non-increasing.
	\label{L3.5}
\end{lemma}

\begin{lemma}
	For system (\ref{2.1})\\
	(i) If there exists $T\geq 0$ such that $u \geq \max_{i \in V}x_{i}(T)$, then for all $t\geq T$, $u\geq \max_{i \in V}x_{i}(t)$, and for $t \geq T$, $\min_{i\in V}x_{i}(t)$ is monotonically non-decreasing.\\
	(ii) If there exists $T\geq 0$ such that $u \leq \min_{i \in V}x_{i}(T)$, then for all $t\geq T$, $u \leq \min_{i \in V}x_{i}(t)$, and for $t \geq T$, $ \max_{i\in V}x_{i}(t)$ is monotonically non-increasing.	
	\label{L3.6}
\end{lemma}

The following two lemmas give the range of opinion of agents in $V_1$ and $V_2$, respectively, when $K_G$ does not contain a cohesive agent set consisting only of unopinionated agents, but there exists a weak cohesive group set $Q^{*}$ consisting only of opinionated agents. 

\begin{lemma}
	For the system (\ref{2.1}), if $K_G$ does not contain a cohesive agent set consisting only of unopinionated agents, but there exists a weak cohesive group set $Q^{*}$ consisting only of opinionated agents, then:\\
	(i) If there exists $T\geq0$ such that for $t\geq T$, $\min_{i \in V}x_{i}(t)$ is monotonically non-decreasing, then for any $i\in V_{2}$, it holds that
	\begin{equation}
		x_{i}(t)\geq \underset{\substack{j\in V_{1}\\t-n_{2}\leq s \leq t-1}}{\min}\,x_{j}(s),\,~\forall\,t\geq T+n_{2}.
	\end{equation}
	(ii) If there exists $T\geq0$ such that for $t\geq T$, $\max_{i \in V}x_{i}(t)$ is monotonically non-increasing, then for any $i\in V_{2}$, it holds that
	\begin{equation}
		x_{i}(t)\leq \underset{\substack{j\in V_{1}\\t-n_{2}\leq s \leq t-1}}{\max}\,x_{j}(s),\,~\forall\,t\geq T+n_{2}.
	\end{equation}
	\label{L3.7}
\end{lemma}

\begin{lemma}
	For system (\ref{2.1}), if $K_G$ does not contain a cohesive agent set consisting only of unopinionated agents, but there exists a weak cohesive group set $Q^{*}$ consisting only of opinionated agents, then\\
	(i) If there exists $T\geq 0$ such that for $t\geq T$, $\min_{i \in V}x_{i}(t)$ is monotonically non-decreasing, then
	\begin{equation}
		x_{i}(t)-u\geq (1-\lambda_{\max})^{K}(\underset{j\in V}{\min}\,x_{j}(T)-u)
		\label{Eq:4.8.1}
	\end{equation}
	for $\forall\, i\in V_{1}\,,~t\geq (K-1)(n_{2}+1)+T+1\,,~K\in \mathbb{Z}^{+}$.
	\vspace{5pt}\\
	(ii) If there exists $T\geq0$ such that for $t\geq T$, $\max_{i \in V}x_{i}(t)$ is monotonically non-increasing, then
	\begin{equation}
		x_{i}(t)-u\leq (1-\lambda_{\min})^{K}(\underset{j\in V}{\max}\,x_{j}(T)-u)
	\end{equation}
	for $\forall\,i\in V_{1}\,,~t\geq (K-1)(n_{2}+1)+T+1\,,~K\in \mathbb{Z}^{+}$.
	\label{L3.8}
\end{lemma}

With the above lemmas as a foundation, we state the main theorem of this section as follows, which provides a sufficient condition for achieving opinion consensus among agents in $K_G$ with partial unopinionated agents.

\begin{theorem}
	System (\ref{2.1}) can achieve asymptotic consensus for any initial opinion $\bm{x}(0)\in \mathbb{R}^{n}$, and the consensus is bias $u$, if $K_G$ does not contain a cohesive agent set consisting only of unopinionated agents, but there exists a weak cohesive group set $Q^{*}$ consisting only of opinionated agents.
	\label{T3.1}
\end{theorem}

\noindent$Proof$:
Consider two cases.\\
$Case$ 1: For $\forall\, t\geq0$, the state of system (\ref{2.1}) satisfies $\min_{i\in V}x_{i}(t) < u < \max_{i\in V}x_{i}(t)$. According to Lemma \ref{L3.5}, we can obtain that for $\forall\, t\geq0$, $\min_{i \in V}x_{i}(t)$ is monotonically non-decreasing, and $\max_{i \in V}x_{i}(t)$ is monotonically non-increasing. Since $\lambda_{i} \in (0,1]$, by Lemma \ref{L3.8}, we have
\begin{equation*}
	\lim_{t \to \infty} x_{i}(t) = u, \,~\forall i \in V_{1}\,,~\bm{x}(0) \in \mathbb{R}^{n}.
\end{equation*}
$Case$ 2: There exists $T \geq 0$ such that $u \leq \min_{i \in V} x_{i}(T)$ or $u \geq \max_{i \in V} x_{i}(T)$. Since these two cases are similar, assume that $u \geq \max_{i \in V} x_i(T)$. By Lemma \ref{L3.6}(1), we have $x_i(t) - u \leq 0$ for all $i \in V_1$ and $t \geq T$, and for all $t \geq T$ and $\min_{i \in V} x_i(t)$ is monotonically non-decreasing, and by Lemma \ref{L3.8}(1), for all opinionated agents, we have
\begin{equation*}
	\lim_{t \to \infty} x_{i}(t) = u, \,~\forall i \in V_{1}\,,~\bm{x}(0) \in \mathbb{R}^{n}.
\end{equation*}
By Lemma \ref{L3.7}, when $t \to \infty$, due to the squeeze theorem, we obtain 
\begin{equation*}
	\lim_{t \to \infty} x_{i}(t) = u, \,~\forall i \in V_{2}\,,~\bm{x}(0) \in \mathbb{R}^{n}.
\end{equation*}
Therefore,we have
\begin{equation*}
	\lim_{t \to \infty} x_{i}(t) = u, \,~\forall i \in V_{1} \cup V_{2}\,,~\bm{x}(0) \in \mathbb{R}^{n}.
\end{equation*}
\endproof

\subsection{Analyzing Fully Opinionated Agents}
\label{B3.4}
Building upon the analysis of opinion convergence with partially opinionated agents in the preceding section, we extend our investigation to the scenario where all agents are inherently opinionated—e.g., social network individuals each holding a fixed core stance on a public issue (e.g., environmental policy) and adjusting their expressed opinions through extrinsic interactions without deviating from their intrinsic positions. Herein, we focus on three core aspects of opinion dynamics within this framework: the analysis of opinion convergence, the quantification of agents’ opinion convergence rate, and the derivation of the analytical expression for the limit point. The opinion update rule for the ``fully opinionated agents'' scenario is formally characterized as follows:
\begin{equation}
	x_{i}(t+1) = \lambda_{i}u_{i}+(1-\lambda_{i})E_{i}(\bm{x}(t))
	\label{4.1}
\end{equation}
for all $i \in V$ and $t \in \mathbb{N}$, let $u_i$ denote the heterogeneity parameter, capturing the agent-specific bias (i.e., the inherent bias varies across different agents). For notational convenience in the proofs of this section, we define:
\begin{equation*}
	Med(\bm{x}(t);\bm{W}):=\left(Med_{1}(\bm{x}(t);\bm{W}), \ldots, Med_{n}(\bm{x}(t);\bm{W})\right)^{\top}.
\end{equation*}
Then (\ref{1.2}) can be rewritten as
\begin{equation}
	E(\bm{x}(t)) \!=\! (\bm{I}_{n}\!-\!\bm{\Gamma})Med(\bm{x}(t);\bm{W})\!+\!\bm{\Gamma} Med(\bm{A}\bm{x}(t);\bm{M})
	\label{E(x)},
\end{equation}
and system (\ref{4.1}) can be rewritten as
\begin{equation}    
	\bm{x}(t+1) = \bm{\Lambda}\bm{u}+(\bm{I}_{n}-\bm{\Lambda})E(\bm{x}(t))
	\label{4}.
\end{equation}

Prior to presenting the key conclusions of this section, we first introduce a set of lemmas that serve as the essential technical underpinnings. These lemmas lay a rigorous foundation for the proofs of the subsequent key conclusions, ensuring the validity and persuasiveness of the derived results.

\begin{lemma}(Non-expansiveness of weighted median mapping\cite{han2024continuous})
	For any $\bm{x},\bm{y}\in\mathbb{R}^{n}$, we have 
	\begin{equation}
		\|Med(\bm{x};\bm{W})-Med(\bm{y};\bm{W})\|_{\infty}\leq\|\bm{x}-\bm{y}\|_{\infty}.
		\label{c}
	\end{equation}
	\label{L4.1}
\end{lemma}

While the non-expansiveness of the weighted median mapping, as documented in existing literature, applies specifically to scenarios where the weight matrix is square, the weight matrix considered herein—one that characterizes environmental influences acting on agents—need not be square. To address this gap, the non-expansiveness of the weighted median mapping for non-square weight matrices is established below.

\begin{corollary}
	For $\forall\, \bm{x},\bm{y}\in\mathbb{R}^{l}$, for a non-square matrix $M=(m_{ij})_{n\times l}$, where $n\neq l$, the inequality in Lemma \ref{L4.1} still holds.
	\label{C4.1}
\end{corollary}

\begin{lemma}(Non-expansiveness of environmental opinion weighted median)
	For $\forall\, \bm{x},\bm{y}\in\mathbb{R}^{n}$, we have
	\begin{equation}
		\|Med(\bm{Ax};\bm{M})-Med(\bm{Ay};\bm{M})\|_{\infty}\leq\|\bm{x}-\bm{y}\|_{\infty}.
	\end{equation} 
	\label{L4.2}
\end{lemma}

Leveraging the lemma established above, the main conclusion of this section is presented below.

\begin{theorem}(Convergence and Convergence Rate) 
	Consider the system (\ref{4}) composed only of opinionated agents, i.e., for all $i \in \bm{V}$, $\lambda_{i} \in (0,1]$ and $u_i \in \mathbb{R}$, then\\
	(i) If the anchoring coefficient $\lambda$ and sensitivity coefficient $\gamma$ satisfy
	\begin{equation}
		(1-\lambda_{\min})\gamma_{\max}<\lambda_{\min}
		\label{t4.1},
	\end{equation}
	then exists a vector $\bm{x}^{*}=(x_{1}^{*},\ldots,x_{n}^{*})^{T}\in \mathbb{R}^{n}$ such that
	\begin{equation*}
		\lim_{t\to \infty}\bm{x}(t)=\bm{x}^{*}
	\end{equation*}
	and the convergence rate is
	\begin{equation*}
		\|\bm{x}(t)-\bm{x}^{*}\|_{\infty}\leq(1-\lambda_{\min})^{t+1} (1+\gamma_{\max})^{t+1}\|\bm{x}(0)-\bm{x}^{*}\|_{\infty}
	\end{equation*} 
	for $\forall\bm{x}(0)\in\mathbb{R}^{n},\, t\in\mathbb{N}^{+}$.
	\vspace{7pt}\\
	(ii) If and only if $u_{1}=u_{2}=,\ldots,=u_{n}=u^{*}$, asymptotic consensus $u^{*}$ can be achieved for any initial opinion.
	\label{T4.1}
\end{theorem}

\noindent$Proof$:
(i) Let
\begin{equation}
	F(\bm{x})=\bm{\Lambda}\bm{u}+(\bm{I}_{n}-\bm{\Lambda})E(\bm{x}),~\forall\, \bm{x}\in\mathbb{R}^{n}
	\label{F(x)}.
\end{equation}
Next, we establish whether $F(\bm{x})$ constitutes a contraction mapping. From (\ref{E(x)}), it follows that
\begin{align}
	&\|F(\bm{x})-F(\bm{y})\|_{\infty} \notag\\
	\leq&\|(\bm{I_{n}-\bm{\Lambda}})\|_{\infty}\|E(\bm{x})-E(\bm{y})\|_{\infty} \notag\\
	\leq&(1-\lambda_{\min})\|\bm{x}-\bm{y}\|_{\infty}(\|\bm{I}_{n}-\bm{\Gamma}\|_{\infty}+\|\bm{\Gamma}\|_{\infty}) \notag\\ 
	\leq&(1-\lambda_{\min})\|\bm{x}-\bm{y}\|_{\infty}(1+\|\bm{\Gamma}\|_{\infty}) \notag\\
	\leq&(1-\lambda_{\min})(1+\gamma_{\max}) \|\bm{x}-\bm{y}\|_{\infty}
	\label{Eq:3.1.1}.
\end{align}
From (\ref{t4.1}), we can further derive that 
\begin{equation*}
	(1-\lambda_{\min})(1+\gamma_{\max})<1.
\end{equation*}
Therefore, $F(\bm{x})$ is a contraction mapping on $\mathbb{R}^{n}$. By the \textit{Banach Fixed-Point Theorem}\cite{banach1922operations}, $F(\cdot)$ admits a unique fixed point $\bm{x}^{*} \in \mathbb{R}^n$ satisfying $F(\bm{x}^{*}) = \bm{x}^{*}$.
From (\ref{4}), we have $\bm{x}(t+1) = F(\bm{x}(t))$. Combining (\ref{Eq:3.1.1}) with the relation $F(\bm{x}^{*}) = \bm{x}^{*}$, we obtain
\begin{align*}
	&\|\bm{x}(t+1)-\bm{x}^{*}\|_{\infty}\notag\\
	=&\|F(\bm{x}(t))-F(\bm{x}^{*})\|_{\infty} \notag\\
	\leq&(1-\lambda_{\min})(1+\gamma_{\max})\|\bm{x}(t)-\bm{x}^{*}\|_{\infty} \notag\\
	\leq&...\leq(1-\lambda_{\min})^{t+1}(1+\gamma_{\max})^{t+1}\|\bm{x}(0)-\bm{x}^{*}\|_{\infty}.
\end{align*}
Since $(1-\lambda_{\min})(1+\gamma_{\max})<1$, when $t\to\infty$, $(1-\lambda_{\min})^{t+1}(1+\gamma_{\max})^{t+1}\to0$, therefore, $\underset{t\to \infty}{\lim}\bm{x}(t)=\bm{x}^{*}$.
\vspace{7pt}\\
(ii)($\Leftarrow$)
When $u_1 = u_2 = \ldots = u_n = u^*$, we show that system (\ref{4}) achieves asymptotic consensus at $u^*$ for arbitrary initial opinions.
\begin{align*}
	F(u^{*}\bm{1}_{n})=
	&\bm{\Lambda}u^{*}\bm{1}_{n}+(\bm{I}_{n}-\bm{\Lambda})E(u^{*}\bm{1}_{n})\\
	=&\bm{\Lambda}u^{*}\bm{1}_{n}+(\bm{I}_{n}-\bm{\Lambda})[(\bm{I}_{n}-\bm{\Gamma})u^{*}\bm{1}_{n}+\bm{\Gamma}u^{*}\bm{1}_{n}]\\
	=&u^{*}\bm{1}_{n}.
\end{align*}
The above results demonstrate that $u^* \bm{1}_n$ is a fixed point of $F(\cdot)$, which implies that system (\ref{4}) can reach consensus at $u^*$.
\vspace{7pt}\\
($\Rightarrow$) 
If system (\ref{4}) asymptotically reaches consensus, and assuming this consensus value is $a^*$, then $a^*$ is known to be a fixed point of $F(\cdot)$; thus,
\begin{align*}
	a^{*}\bm{1}_{n}=&F(a^{*}\bm{1}_n)\\
	=&\bm{\Lambda}\bm{u}+(\bm{I}_n-\bm{\Lambda})[(\bm{I}_n-\bm{\Gamma)}a^{*}\bm{1}_n+\bm{\Gamma}\bm{A}a^{*}\bm{1}_n]\\
	=&\bm{\Lambda}\bm{u}+a^{*}\bm{1}_n-\bm{\Lambda}a^{*}\bm{1}_n.
\end{align*}
The above results lead to the conclusion that $\bm{\Lambda}\bm{u} = \bm{\Lambda} a^* \bm{1}_n$. Given that $\bm{\Lambda}$ is invertible, it follows that $\bm{u} = a^* \bm{1}_n$.
\endproof

Owing to the nonlinearity of the weighted median mechanism and the lack of an analytical expression for it, obtaining an analytical solution to the fixed point of the corresponding contraction mapping poses significant challenges; consequently, an analytical expression for the limit point a remains elusive. Nevertheless, by leveraging the inherent properties of the weighted median, we can establish a  mathematical characterization of the limit point $x^*$.

\begin{definition}(Indicator Function)
	For any subset $B \subseteq V$, define the indicator function for agent $i$,
	\begin{equation}
		\mathbb{I}_{B}(i):= 
		\begin{cases}1, & \text{if}~i \in B; \\ 0, & \text{otherwise}. \end{cases}
	\end{equation}
	\label{D4.1}
\end{definition}

Using the definition of the indicator function, we introduce two descriptive matrices $\bm{P}$ and $\bm{Q}$.

By the definition of the weighted median, for any $i \in V$, the value $\text{Med}_i(\bm{x}; \bm{W})$ is a component of the vector $\bm{x}$. This implies that there exists an agent $k_i \in V$ such that $\text{Med}_i(\bm{x}; \bm{W}) = x_{k_i}$.

Next, we introduce two descriptive matrices $\bm{P}$ and $\bm{Q}$.
\begin{equation*}
	\bm{Px}=(\bm{I}_{n}-\bm{\Lambda})(\bm{I}_{n}-\bm{\Gamma})Med(\bm{x};\bm{W}),~\forall \bm{x} \in \mathbb{R}^{n}.
\end{equation*}
where $p_{ij} = (1-\lambda_i)(1-\gamma_i) \mathbb{I}_{k_i}(j)$. Here, $k_i$ denotes the index of the non-zero entry in the $i$-th row of $\bm{P}$, with its selection depending on the $i$-th row of $\bm{W}$. It is straightforward to verify that the $i$-th row contains exactly one non-zero entry, specifically $(1-\lambda_i)(1-\gamma_i)$.
\begin{equation*}
	\bm{Q(Ax)}=(\bm{I}_{n}-\bm{\Lambda})\bm{\Gamma}Med(\bm{Ax};\bm{M}),~\forall \bm{x} \in \mathbb{R}^{n}.
\end{equation*}
where $q_{ij} = (1-\lambda_i)\gamma_i \mathbb{I}_{l_i}(j)$. Here, $l_i$ denotes the index of the non-zero entry in the $i$-th row of $\bm{Q}$, with its selection depending on the $i$-th row of $\bm{M}$. It is straightforward to verify that the $i$-th row contains exactly one non-zero entry, specifically $(1-\lambda_i)\gamma_i$.

Using the descriptive matrices $\bm{P}$ and $\bm{Q}$, system (\ref{4}) can be rewritten as
\begin{equation}    
	\bm{x}(t+1) = \bm{\Lambda}\bm{u}+\bm{Px}(t)+\bm{Q}(\bm{Ax}(t))
	\label{PQ}.
\end{equation}

\begin{corollary}
	Consider the system (\ref{PQ}), the expression of the limit point is
	\begin{equation}
		\bm{x^{*}}=(\bm{I}_n-\bm{P}-\bm{QA})^{-1}\bm{\Lambda}\bm{u}
		\label{c2},
	\end{equation}
	where $\bm{P}$ and $\bm{Q}$ are newly defined descriptive matrices.
	\label{C4.2}
\end{corollary}

\begin{lemma}
	$\bm{I}_n-\bm{P}-\bm{QA}$ is an invertible matrix.
	\label{L4.3}
\end{lemma}

\section{Simulations}
\label{B4}
\begin{figure}[htbp]
	\centering
	\includegraphics[width = 0.48\textwidth]{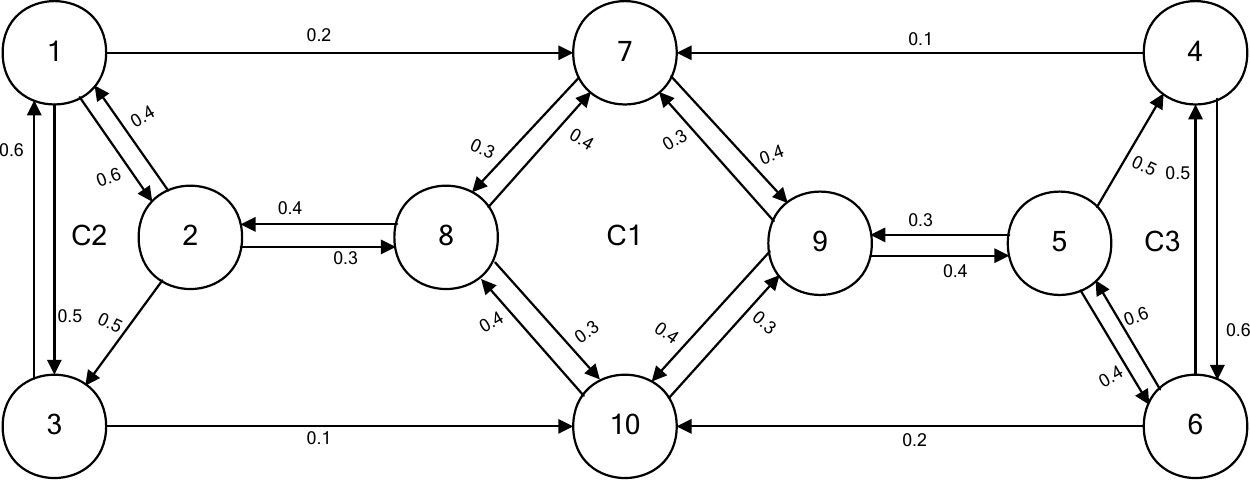}
	\caption{The visualization example presents the simplicial complex considered in the simulation part, which includes $10$ agents and $3$ simplices: $C_{1}$ is a $3$-simplex, $C_{2}$ and $C_{3}$ are $2$-simplex, where $\delta_{C_{1}}=\{7, 8, 9, 10\}$, $\delta_{C_{2}}=\{1, 2, 3\}$ and $\delta_{C_{3}}=\{4, 5, 6\}$. The existence of an arrow between two nodes in the graph indicates that one agent will affect the other agent, and the number near the arrow represents the influence agent weight. In addition, this simplicial complex ignores the internal connections within the $C_{1}$ simplex for convenience of drawing.}
	\label{fig1}
\end{figure}

This section considers the system in Fig.\ref{fig1} and two scenarios under this system, respectively: one is a heterogeneous system that includes both opinionated agents and unopinionated agents, as shown in Fig.\ref{fig_a}(a); the other is a homogeneous system that only includes opinionated agents, as shown in Fig.\ref{fig_a}(b). Furthermore, since the theories in Section \ref{B3} and Section \ref{B4} hold for any initial opinion. Without loss of generality, we assign the initial opinions of $10$ agents as $(-0.4, -0.3, -0.2, -0.1, 0, 0.1, 0.2, 0.3, 0.4, 0.5)^{\top}$ in both situations.

To rule out the interference of possible coupling between initial opinion values and the model structure on the experimental results, a dedicated validation is presented in the \hyperlink{app:L}{Appendix L}.
\subsection{Heterogeneous System}
\begin{figure}[htbp]
	\centering
	\subfloat[]{\centering\includegraphics[width=0.48\textwidth]{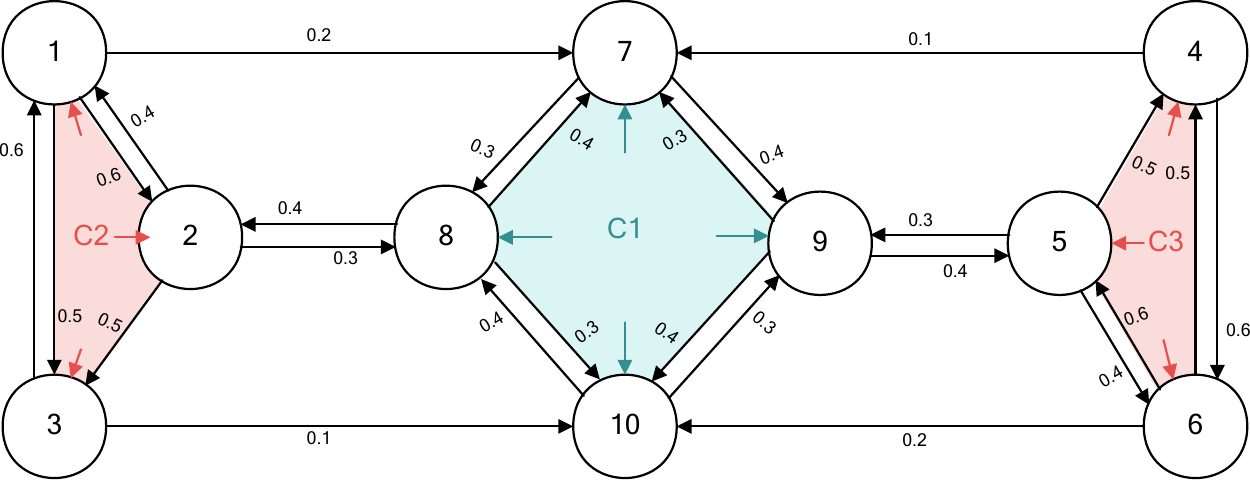} \label{fig5}}
	
	\vspace{10pt}
	
	\subfloat[]{\centering\includegraphics[width=0.48\textwidth]{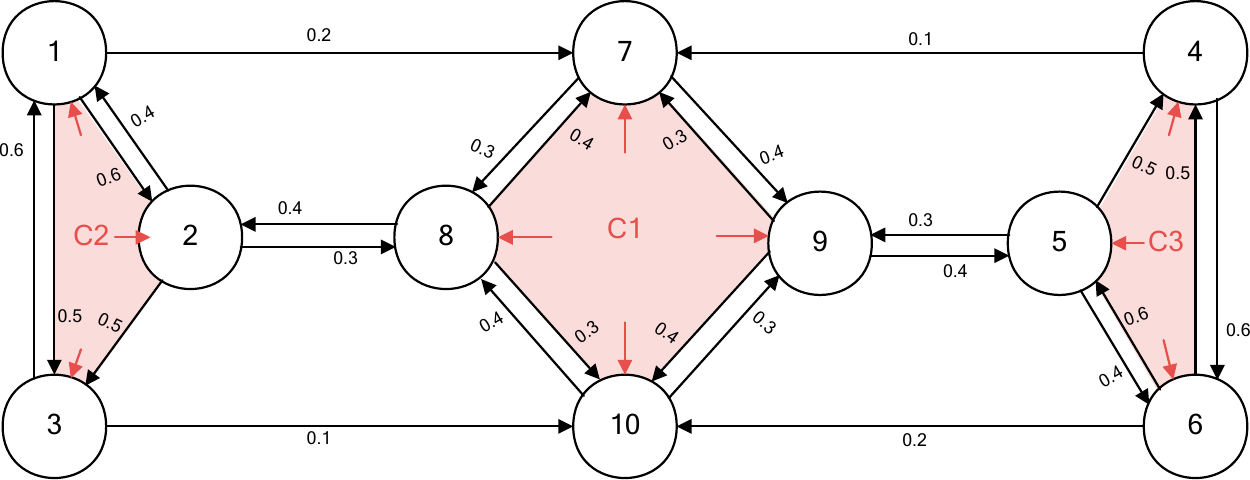} \label{fig2}}
	\caption{
		Simplicial complexes of the simulation example of different system. There are three simplexes in the system, which are $\delta_{C_{1}}=\{7, 8, 9, 10\}$, $\delta_{C_{2}}=\{1, 2, 3\}$ and $\delta_{C_{3}}=\{4, 5, 6\}$. (a) Heterogeneous system: $\delta_{C_{1}}$ is a simplex composed of unopinionated agents, while $\delta_{C_{2}}$ and $\delta_{C_{3}}$ are simplices composed of opinionated agents. Furthermore, according to the weight of agents, $C_{1}$ is a cohesive agent set composed of unopinionated agents. (b) Homogeneous system: $\delta_{C_{1}}$, $\delta_{C_{2}}$ and $\delta_{C_{3}}$ are simplices composed of opinionated agent. In both systems, the following notations and rules apply consistently: Black arrows represent the influence between agents, and colored arrows represent the influence of simplices on agent. The numbers near the arrows represent the influence weights. If no number is marked, the influence weight is 1. Nodes in the red-covered area represent agents with bias, while those in the green-covered area represent unopinionated agents.}
	\label{fig_a}
\end{figure}

As shown in Fig.\ref{fig_a}(a), we assume that agent $\{1,\ldots,6\}$ in the system are opinionated agents, and we set each bias $u_{i}=0$ and uniformly randomly select the anchoring coefficient $\lambda_{i}$ within $(0, 1]$, while $\{7,\ldots,10\}$ are unopinionated agent, with anchoring coefficient $\lambda_{i}=0$. Additionally, regardless of whether they are opinionated agent or unopinionated agent, the sensitivity coefficient $\gamma_{i}$ is uniformly randomly selected within $(0, 1]$.

We observe that in Fig.\ref{fig_a}(a), there exists a cohesive agent set formed by unopinionated agents $\{7,\ldots,10\}$, and no weak cohesive group set formed by simplices. From the conclusions of this work, it can basically be inferred that this system will not form a consensus opinion. Indeed, after simulation experiments in Fig.\ref{python_a}(a), we found that the system eventually formed two opinion stable states. Opinionated agents form subgroups and take the bias value as their consensus opinion. Unopinionated agents attract each other, and their different opinions converge towards each other, deviating from the consensus opinion of opinionated agents.

In order to enable the system to asymptotically reach a consensus, we slightly adjust the weights in the system to disrupt the cohesive agent set composed of unopinionated agents and form a weak cohesive group set composed of opinionated agents. The specific operation is as follows.

\begin{figure}[htbp]
	\centering
	\subfloat[]{\centering\includegraphics[width=0.48\textwidth]{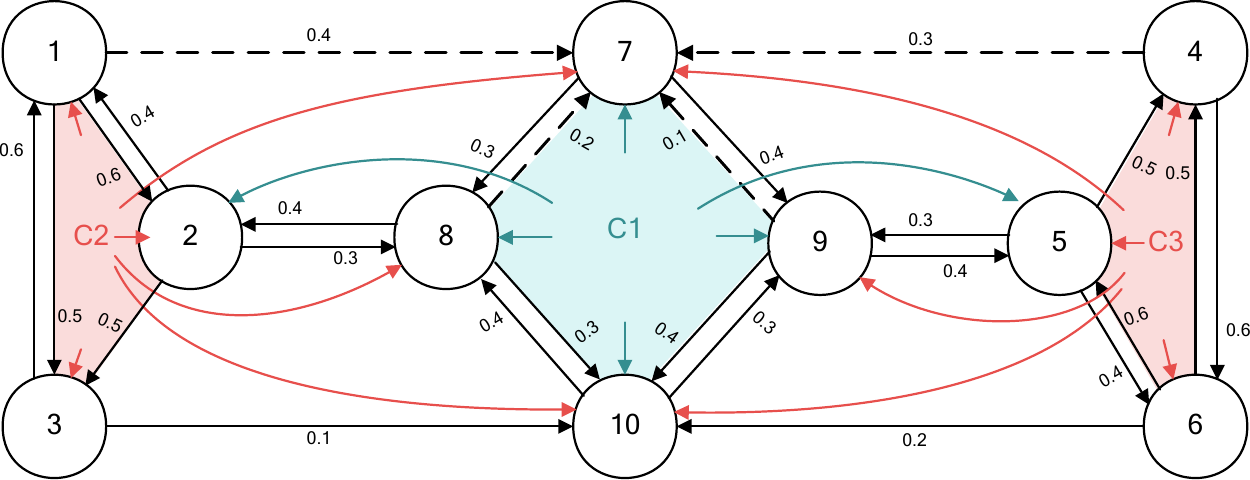} \label{fig3}}
	
	\vspace{10pt}
	
	\subfloat[]{\centering\includegraphics[width=0.48\textwidth]{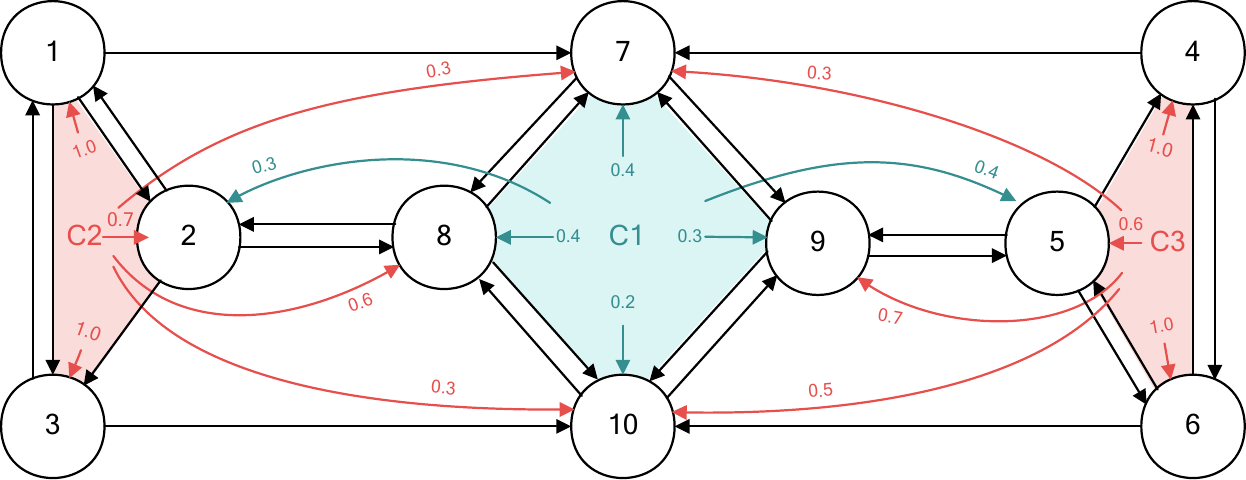} \label{fig4}}
	
	\caption{Illustration of influence weights in different dimensions. To make Fig.\ref{fig_a}(a) meet the conditions of Theorem \ref{T3.1}, the low-order and high-order influence weights are modified. (a) Influence weight between agents: By adjusting the weights between agent 7 and its neighbors, where dashed arrows represent the adjusted weights, the cohesive agent set $C_{1}$ composed of unopinionated agents in Fig.\ref{fig_a}(a) is disrupted. (b) Influence weight of the environment on agents: By changing the influence weights of the simplex on agents, the system forms a weak cohesive group set $\{C_{2}, C_{3}\}$ composed of opinionated agents. 
	}
	\label{fig_b}
\end{figure}

\begin{figure}[htbp]
	\centering
	\subfloat[]{\includegraphics[width=0.23\textwidth]{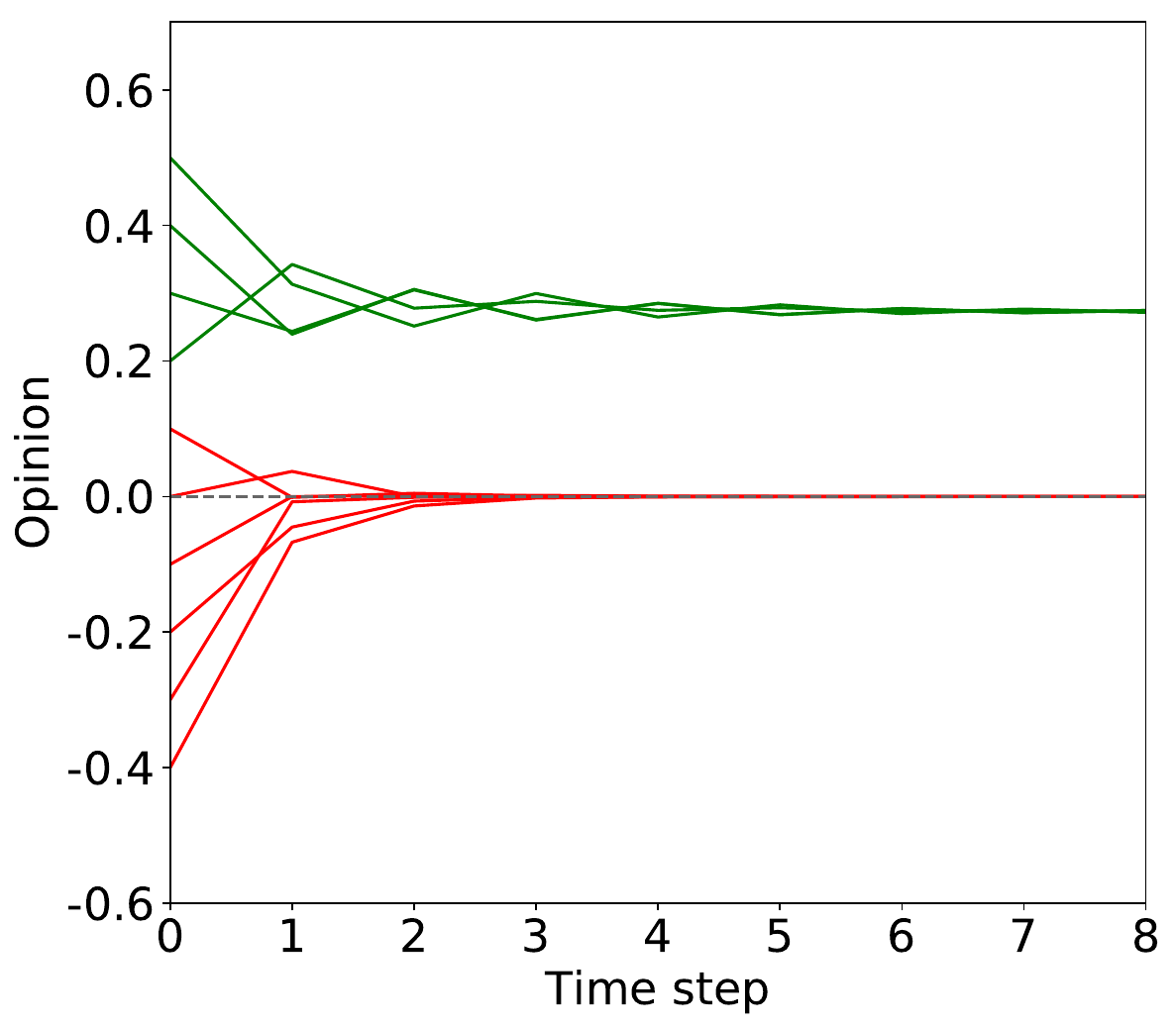} \label{python1}}
	\hfill
	\subfloat[]{\includegraphics[width=0.23\textwidth]{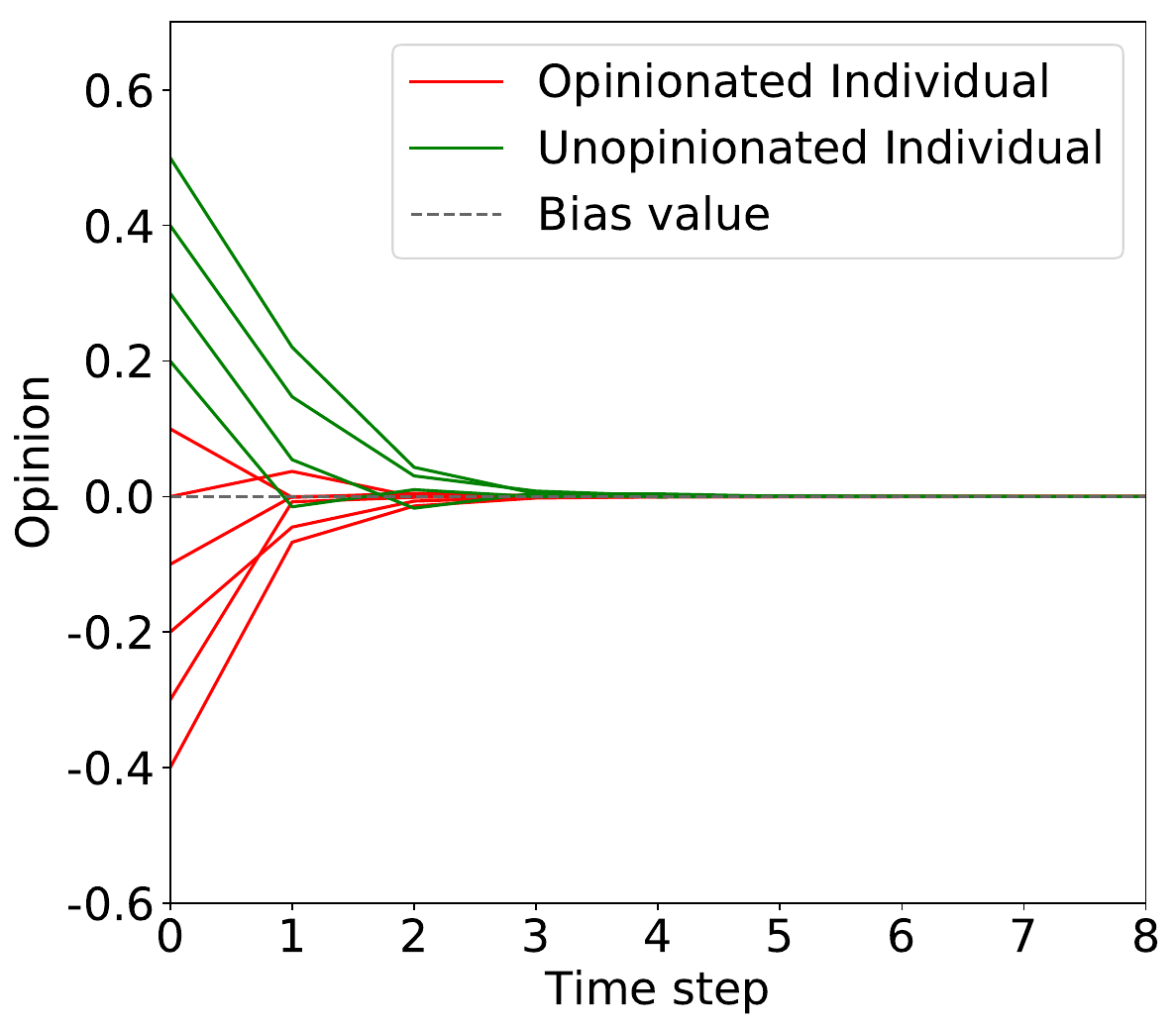} \label{python2}}
	\caption{The evolution processes of opinion for heterogeneous system, starting from different initial opinions of agents at time 0. Red represents opinionated agents with the same bias, green represents unopinionated agents, and the dashed line represents the common bias of all opinionated agents.
		(a) Before weight adjustment, the system cannot reach a consensus, but instead forms two stable states. 
		(b) After the weight adjustment, the system can reach a consensus, which does not contain a cohesive agent set composed of unopinionated agents and contain a weak cohesive group set composed of opinionated agents.}
	\label{python_a}
\end{figure}

First, interchange the weights $w_{17}$ and $w_{87}$, and interchange the weights $w_{47}$ and $w_{97}$ in Fig.\ref{fig_b}(a). This disrupts the original cohesive agent set composed of unopinionated agents $\{7,\ldots,10\}$. Consequently, there is no cohesive agent set composed of unopinionated agents in the system. According to Theorem \ref{T3.1}, this satisfies one condition for the system to progressively reach consensus. 

Next, adjust the weights of the simplices for each agent to form a weak cohesive group of opinionated agents in Fig.\ref{fig_b}(b). This satisfies another condition for the system to asymptotically reach consensus.

At this point, the system in Fig.\ref{fig_b} with adjusted weights satisfies the conditions of Theorem \ref{T3.1}, so we can conclude that the system will certainly achieve asymptotic consensus, and the consensus value is the bias of the opinionated agents. Indeed, through simulation experiments in Fig.\ref{python_a}(b), we discovered that the opinions of unopinionated agents in the system no longer deviate, and all agents reach a unified consensus with a consensus value of 0.

\subsection{Homogeneous System}
\begin{figure}[htbp]
	\centering
	\subfloat[]{\includegraphics[width=0.23\textwidth]{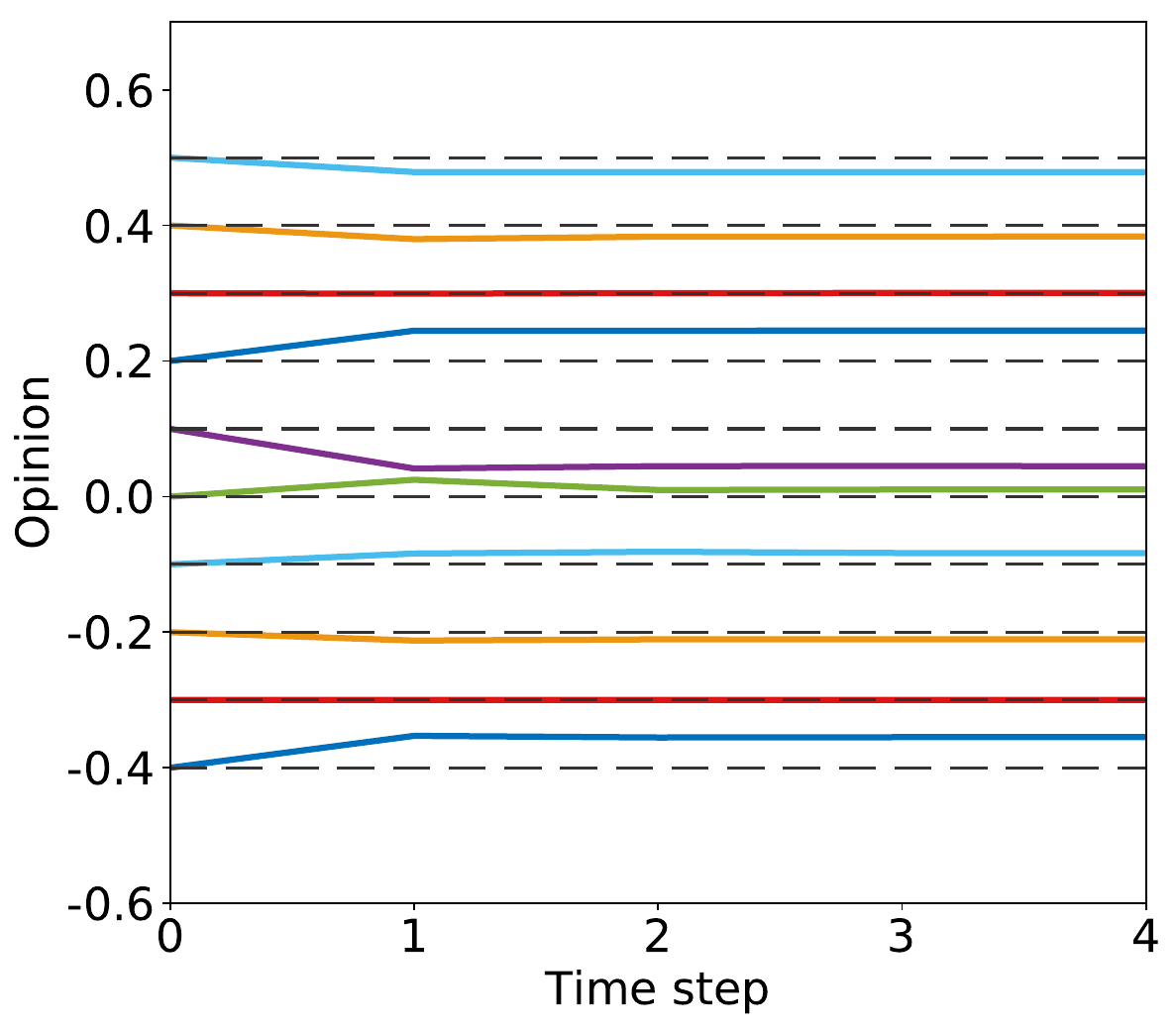} \label{python3}}
	\hfill
	\subfloat[]{\includegraphics[width=0.23\textwidth]{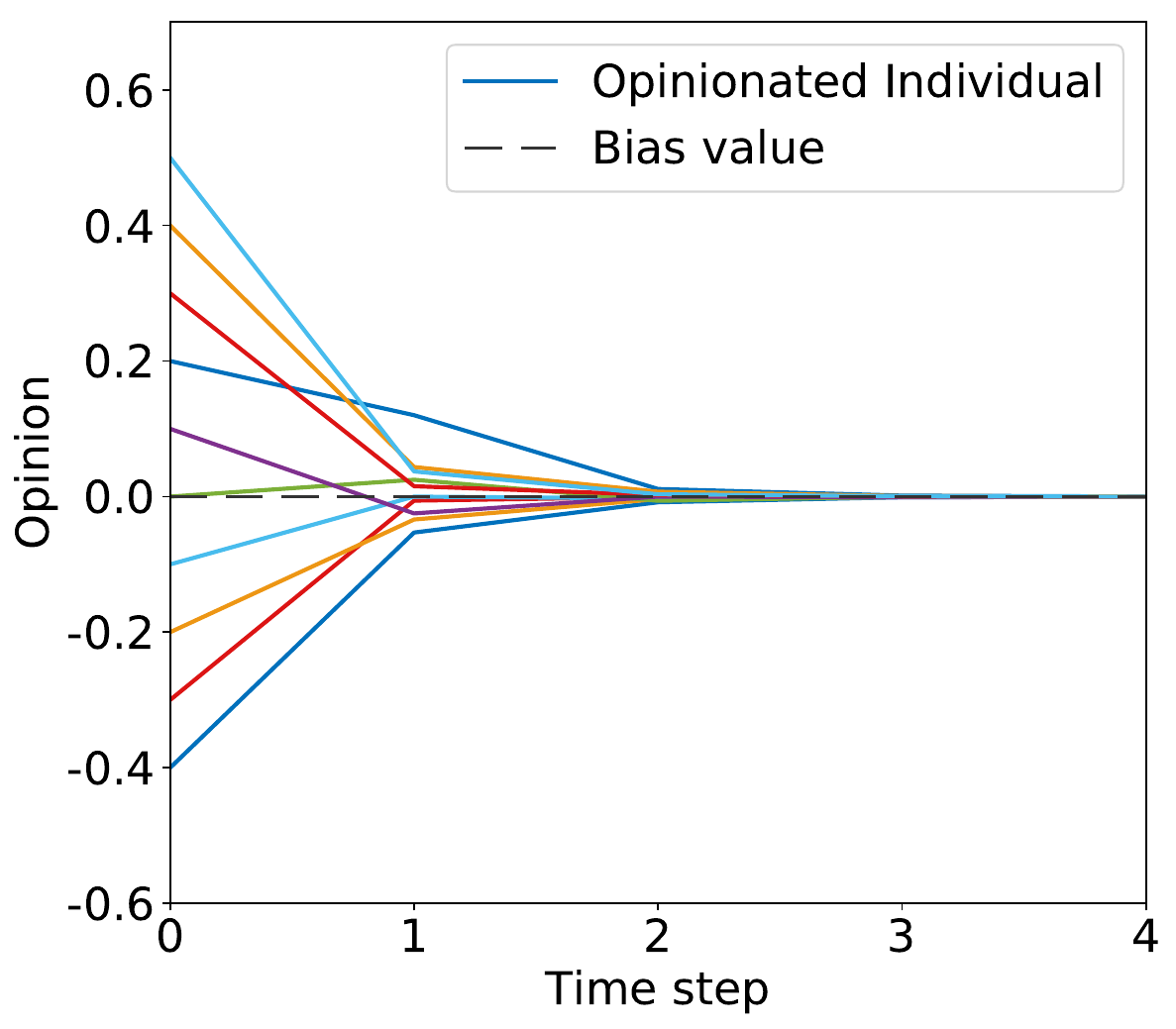} \label{python4}}
	\caption{The evolution processes of opinion for homogeneous system, starting from diverse initial opinions of agents at time 0. All agents in this system are opinionated, with different colors representing distinct opinionated agents.
		(a) Each agent possesses a different bias, and dashed lines represent the bias values of agents. 
		(b) Each agent shares the same bias value.
	}
	\label{python_b}
\end{figure}

In Fig.\ref{fig_a}(b), all agent in the system are opinionated, and the bias $u_{i}$ is set to the initial opinion $x_{i}(0)$. Under the condition that the inequality (\ref{t4.1}) is satisfied, the anchoring coefficient $\lambda_{i}$ and the sensitive coefficient $\gamma_{i}$ are randomly and uniformly selected within $(0, 1]$. According to Theorem \ref{T4.1}(1), the system will tend towards a stable state where opinion converges. Indeed, after simulation experiments, we found the system eventually converges to a stable state, where each agent holds its own distinct stable opinion in Fig.\ref{python_b}(a). Additionally, if we set the bias $u_{i}$ to the same value, assuming they are all set to $0$, according to Theorem \ref{T4.1}(2), the system will not only tend towards a stable state where opinion converges, but also reach a consensus, and the consensus value will be the same bias 0. Indeed, the simulation experiment confirmed our conclusion in Fig.\ref{python_b}(b).

The inequality condition (\ref{t4.1}) in Theorem \ref{T4.1}(1) can be understood in light of real-world situations as follows: In a system, the degree to which the most susceptible agents rely on the environment or their neighbors must be less than the degree to which the least stubborn agents adhere to their own biases. Specifically, an agent’s external dependence cannot exceed their intrinsic anchoring.
In real life, when every agent has their own biases, only if agents maintain a strong anchoring to their own biases and are not easily influenced by the environment, their opinions will eventually remain stable despite minor fluctuations. If all agents have the same biases, naturally, their opinions will eventually reach a consensus on these biases.

\section{Conclusion}\label{B5}
This work introduces high-order influence into the weighted median opinion dynamics model, ingeniously incorporating it into the opinion evolution process by constructing the model on a simplicial complex, and conducts a theoretical analysis of this dynamic behavior. Firstly, for a heterogeneous system consisting of both opinionated and unopinionated agents, we provide sufficient conditions for the system to asymptotically reach consensus, and extract special structures related to the evolution of opinions in the system based on the structure of simplicial complexes. Additionally, for a system composed entirely of opinionated agents, we present the convergence and convergence rate of the system. The verification through simulation experiments provides a good practical explanation for the theoretical analysis.

There are still some issues to be addressed in the future. For instance, at present, Theorem \ref{T3.1} only serves as a sufficient condition. In subsequent research, it remains to be seen whether a necessary and sufficient condition for the asymptotic consensus of the system under the influence of higher-order simplicial complexes can be obtained. Another interesting question is that system (\ref{2.1}) only considers the situation where all opinionated agents have the same bias. However, if they have different biases, the convergence of the system (\ref{2.1}) remains unknown. Additionally, if the system contains weak cohesive groups composed of agents with different biases, it is necessary to explore whether multiple opinion domains can be formed. We leave these questions for future work.

\section*{Appendix}

\subsection{Proof of Lemma \ref{L3.1}}
\noindent Given $\bm{y} = \bm{A}\bm{x}$ with $\bm{y} = (y_1, y_2, \dots, y_l)^T$, we first define the minimum and maximum entries of $\bm{y}$ as:
\begin{equation}
	y_{\alpha}=\underset{j \in \{1,2,..,l\}}{\min}\,y_{j},~~
	y_{\beta}=\underset{j \in \{1,2,..,l\}}{\max}\,y_{j}\label{Eq:4.1.1}.
\end{equation}
where $\alpha, \beta \in \{1, 2, \dots, l\}$ denote the indices corresponding to the minimum and maximum entries of $\bm{y}$, respectively. Leveraging the matrix-vector multiplication $\bm{y} = \bm{A}\bm{x}$, the entries $y_{\alpha}$ and $y_{\beta}$ admit explicit expressions as:
\begin{equation}     y_{\alpha}=\bm{a}_{\alpha}\bm{x},~~y_{\beta}=\bm{a}_{\beta}\bm{x}\label{Eq:4.1.2}.
\end{equation}
where $\bm{a}_{\alpha}$ and $\bm{a}_{\beta}$ denote the $\alpha$-th and $\beta$-th row vectors of matrix $\bm{A}$, respectively.
From (\ref{Eq:4.1.1})-(\ref{Eq:4.1.2}) and the matrix $\bm{A}$ is a row stochastic matrix, we obtain
\begin{align*}
	&\bm{a}_{\alpha}\bm{x}
	\leq Med_{i}(\bm{Ax};\bm{M})
	\leq \bm{a}_{\beta}\bm{x}\\
	\Rightarrow 
	&\bm{a}_{\alpha}[(\underset{i \in V}{\min}\,x_{i})\boldsymbol{1}_{n}]
	\leq Med_{i}(\bm{Ax};\bm{M})
	\leq \bm{a}_{\beta}[(\underset{i \in V}{\max}\,x_{i})\boldsymbol{1}_{n}]\\
	\Rightarrow
	&\underset{i \in V}{\min}\,x_{i}
	\leq Med_{i}(\bm{Ax};\bm{M})
	\leq \underset{i \in V}{\max}\,x_{i}.
\end{align*}
Therefore, for $\forall i \in V$ and $\forall \bm{x} \in \mathbb{R}^{n}$, the inequality $\min_{j \in V}x_{j}
\leq Med_{i}(\bm{Ax};\bm{M})
\leq \max_{j \in V}x_{j}$ holds.

\subsection{Proof of Lemma \ref{L3.3}}
\noindent $P$ is a cohesive influential cluster. According to Definition \ref{D3.4}, $P$ is a cohesive agent set and is associated with a strong cohesive group set $Q$. Firstly, since the $P$ is a cohesive agent set, according to Lemma \ref{L3.2} and Definition \ref{D3.1}, for $\forall\,i \in P$, we have
\begin{equation}
	\underset{j\in P}{\min}\,x_{j}\leq Med_{i}(\bm{x};\bm{W})\leq 
	\underset{j\in P}{\max}\,x_{j} \label{Eq:4.2.-1}.
\end{equation}
Next, we prove that $\underset{j\in P}{\min}\,x_{j}\leq Med_{i}(\bm{Ax};\bm{M})\leq 
\underset{j\in P}{\max}\,x_{j}$.
Given $\bm{y}=\bm{Ax}$, we have
\begin{equation}
	Med_{i}(\bm{Ax};\bm{M})=Med_{i}(\bm{y};\bm{M})\label{Eq:4.2.0}.
\end{equation}
Let $y_{s_{1}},y_{s_{2}},\ldots,y_{s_{l}}$ is a reordering of $y_{1},y_{2},\ldots,y_{l}$, such that
\begin{equation}
	y_{s_{1}}\leq y_{s_{2}}\leq\ldots\leq y_{s_{l}} \label{Eq:4.2.1}.
\end{equation}
Define $\alpha=\min\{t\in\{1,...,l\}:s_{t}\in Q\}$ and $\beta=\max\{t\in\{1,...,l\}:s_{t}\in Q\}$. Leveraging (\ref{Eq:4.2.1}), we obtain
\begin{equation}
	Q\subseteq\{s_{\alpha},...,s_{\beta}\}
	\label{Eq:4.2.2}
\end{equation}
and
\begin{equation}
	y_{s_{\alpha}} = \underset{k\in Q}{\min}\,y_{k},\quad y_{s_{\beta}} = \underset{k \in Q}{\max}\,y_{k}
	\label{Eq:4.2.3}.
\end{equation}
Since $Q$ is a strong cohesive group set, for any $i \in P \subseteq V$, it follows that $\sum_{k\in Q}m_{ik}>\frac{1}{2}$. Combining this with (\ref{Eq:4.2.2}), we deduce $\sum_{t = \alpha}^{\beta}m_{is_{t}}\geq \sum_{k\in Q}m_{ik}>\frac{1}{2}$. Building on this finding, we can further derive
\begin{equation}
	\sum_{t \geq \alpha}m_{is_{t}} > \frac{1}{2}, ~~\sum_{t\leq \beta}m_{is_{t}} > \frac{1}{2}
	\label{Eq:4.2.4}.
\end{equation}
Let $y^{*} = Med_{i}(\bm{y};\bm{M})$. \\
If $y^{*} < y_{s_{\alpha}}$, then from (\ref{Eq:4.2.4}) we can get $\sum_{k:\,y_{k}>y^{*}}m_{ik} > \frac{1}{2}$, which contradicts the definition of weighted median. \\
If $y^{*} >  y_{s_{\beta}}$, then from (\ref{Eq:4.2.4}), we can get $\sum_{k:\,y_{k}<y^{*}}m_{ik} > \frac{1}{2}$, which contradicts the definition of weighted median. \\
Therefore, $y_{s_{\alpha}}\leq y^{*}\leq y_{s_{\beta}}$. From (\ref{Eq:4.2.3}), we know that
\begin{equation}
	\underset{k\in Q}{\min}\,y_{k}\leq Med_{i}(\bm{y};\bm{M})\leq \underset{k\in Q}{\max}\,y_{k}\label{Eq:4.2.5}.
\end{equation}
Given that $Q$ denotes a strong cohesive group set, every agent within simplex $k$ belongs to $P$. Let $\delta_{k}$ represent the set of agents constituting simplex $k$; it then follows that
\begin{equation}
	\delta_{k}\subset P, \,~\forall\, k\in Q \label{Eq:4.2.6}.
\end{equation}
By virtue of $\bm{y}=\bm{Ax}$, the $k$-th entry of $\bm{y}$ satisfies
\begin{equation}
	y_{k}=\bm{a}_{k}\bm{x},\,~\forall\, k\in Q \label{Eq:4.2.7},
\end{equation}
with $\bm{a}_{k}$ denoting the $k$-th row vector of matrix $\bm{A}$. Combining (\ref{Eq:4.2.6}) and the definition of matrix $\bm{A}$, we derive
\begin{equation*}
	\bm{a}_{k}\bm{x}
	\geq \bm{a}_{k}[(\underset{i\in \delta_{k}}{\min}\,x_{i})\boldsymbol{1}_{n}]
	\geq \bm{a}_{k}[(\underset{i\in P}{\min}\,x_{i})\boldsymbol{1}_{n}]
	= \underset{i\in P}{\min}\,x_{i}[\bm{a}_{k}\boldsymbol{1}_{n}],
\end{equation*}
\begin{equation}
	\bm{a}_{k}\bm{x}
	\leq\bm{a}_{k}[(\underset{i\in \delta_{k}}{\max}\,x_{i})\boldsymbol{1}_{n}]
	\leq\bm{a}_{k}[(\underset{i\in P}{\max}\,x_{i})\boldsymbol{1}_{n}]
	= \underset{i\in P}{\max}\,x_{i}[\bm{a}_{k}\boldsymbol{1}_{n}]
	\label{Eq:4.2.8}.
\end{equation}
Since the matrix $\bm{A}$ is a row stochastic matrix and (\ref{Eq:4.2.7}), the above (\ref{Eq:4.2.8}) can be further derived as
\begin{equation}
	\underset{i \in P}{\min}\,x_{i} 
	\leq y_{k}
	\leq \underset{i \in P}{\max}\,x_{i},\,~\forall\, k\in Q.
\end{equation}
Further, we can obtain
\begin{equation*}
	\underset{i \in P}{\min}\,x_{i}
	\leq \underset{k \in Q}{\min}\,y_{k}
	\leq \underset{k \in Q}{\max}\,y_{k}
	\leq \underset{i \in P}{\max}\,x_{i}.
\end{equation*}
From (\ref{Eq:4.2.5}) and (\ref{Eq:4.2.0}), we can obtain
\begin{equation}
	\underset{i \in P}{\min}\,x_{i}
	\leq Med_{i}(\bm{Ax};\bm{M})
	\leq \underset{i \in P}{\max}\,x_{i} \label{Eq:4.2.9}.
\end{equation}
From (\ref{Eq:4.2.-1}), (\ref{Eq:4.2.9}) and (\ref{1.2}) we can obtain
\begin{equation*}
	\underset{j\in P}{\min}\,x_{j}
	\leq 
	E_{i}(\bm{x})
	\leq \underset{j\in P}{\max}\,x_{j}.
\end{equation*} 
Therefore, we have (\ref{3.3}) hold. This completes the
proof of this lemma.

\subsection{Proof of Lemma \ref{L3.4}}
\noindent Since all agents in $P$ are unopinionated, from (\ref{2.1}), for $\forall\, i \in P,~\, t \in \mathbb{N}$, we have
\begin{equation}
	x_{i}(t+1)=E_{i}(\bm{x}(t))
	\label{Eq:4.4.1}.
\end{equation}
Since $P$ is a cohesive influential cluster, according to Lemma \ref{L3.3}, for $\forall\, i \in P,~\, t \in \mathbb{N}$, we have
\begin{equation*}
	\underset{j \in P}{\min}\,x_{j}(t)
	\leq E_{i}(\bm{x}(t))
	\leq \underset{j \in P}{\max}\,x_{j}(t).
\end{equation*}
According to (\ref{Eq:4.4.1}), for $\forall\, i \in P,~\, t \in \mathbb{N}$, we can obtain
\begin{equation*}
	\underset{j \in P}{\min}\,x_{j}(t)
	\leq x_{i}(t+1)
	\leq \underset{j \in P}{\max}\,x_{j}(t).
\end{equation*}
Further, for $\forall\, t \in \mathbb{N}$, we can obtain
\begin{equation}
	\underset{j \in P}{\min}\,x_{j}(t)
	\leq \underset{i \in P}{\min}\,x_{i}(t+1)
	\leq \underset{i \in P}{\max}\,x_{i}(t+1)
	\leq \underset{j \in P}{\max}\,x_{j}(t)
	\label{3.1}.
\end{equation}
By repeatedly using (\ref{3.1}), for $\forall\, t \in \mathbb{N}$, we can obtain
\begin{equation*}
	\underset{j \in P}{\min}\,x_{j}(0)
	\leq \underset{i \in P}{\min}\,x_{i}(t)
	\leq \underset{i \in P}{\max}\,x_{i}(t)
	\leq \underset{j \in P}{\max}\,x_{j}(0).
\end{equation*}
That is
\begin{equation*}
	\underset{j \in P}{\min}\,x_{j}(0)
	\leq x_{i}(t)
	\leq \underset{j \in P}{\max}\,x_{j}(0),
	\,~\forall i \in P,\, t \in \mathbb{N}.
\end{equation*}
The proof is complete.		

\subsection{Proof of Lemma \ref{L3.5}}
\noindent(i) For $\forall\, i \in V_{2}$, according to (\ref{2.1}), (\ref{1.2}) and Lemma \ref{L3.1}, we have
\begin{align*}
	x_{i}(t+1)& = E_{i}(\bm{x}(t))\\
	& = (1-\gamma_{i})\,Med_{i}(\bm{x}(t);\bm{W})+\gamma_{i}\,Med_{i}(\bm{Ax}(t);\bm{M})\\
	&\geq \underset{j \in V}{\min}\,x_{j}(t),
	~\forall \, t\geq T.
\end{align*}
Furthermore, we can obtain
\begin{equation}
	\underset{i \in V_{2}}{\min}\,x_{i}(t+1)\geq \underset{j \in V}{\min}\,x_{j}(t),
	~\forall\,t\geq T.
	\label{Eq:4.5.1}
\end{equation}
For $\forall\, i \in V_{1}$, according to (\ref{2.1}), Lemma \ref{L3.1} and the known $u \geq \min_{i\in V}\,x_{i}(t)$, we have
\begin{align*}
	x_{i}(t+1)&=\lambda_{i}u+(1-\lambda_{i})E_{i}(\bm{x}(t))\\
	&\geq \underset{j\in V}{\min}\,x_{j}(t),
	~\forall\,t\geq T.
\end{align*}
Furthermore, we can get
\begin{equation}
	\underset{i\in V_{1}}{\min}\,x_{i}(t+1)\geq \underset{j\in V}{\min}\,x_{j}(t),
	~\forall\,t\geq T.
	\label{Eq:4.5.2}
\end{equation}
From (\ref{Eq:4.5.1}) and (\ref{Eq:4.5.2}), it can be deduced that
\begin{equation*}
	\underset{i\in V}{\min}\,x_{i}(t+1)\geq \underset{j\in V}{\min}\,x_{j}(t),
	~\forall\,t\geq T.
\end{equation*}
That is, $ \min_{i\in V}x_{i}(t)$ is monotonically non-decreasing.\\
(ii) The proof is similar to (i).

\subsection{Proof of Lemma \ref{L3.6}}
\noindent(i) For $\forall\, i \in V_{2}$, according to (\ref{2.1}), (\ref{1.2}), Lemma \ref{L3.1} and the known $u \geq \max_{i \in V}x_{i}(T)$, we have
\begin{align*}
	x_{i}(T+1)& = E_{i}(\bm{x}(T))\\
	& = (1-\gamma_{i})\,Med_{i}(\bm{x}(T);\bm{W})+\gamma_{i}\,Med_{i}(\bm{Ax}(T);\bm{M})\\
	&\leq\underset{j \in V}{\max}\,x_{j}(T)\\
	&\leq u.
\end{align*}
Furthermore, we can get
\begin{equation}
	\underset{i \in V_{2}}{\max}\,x_{j}(T+1)\leq u.
	\label{Eq:4.6.1}
\end{equation}
For $\forall\, i \in V_{1}$, according to (\ref{2.1}), Lemma \ref{L3.1} and the known $u \geq \max_{i \in V}x_{i}(T)$, we have $x_{i}(T+1) = \lambda_{i}u+(1-\lambda_{i})E_{i}(\bm{x}(t))\leq u$.
It then follows that
\begin{equation}
	\underset{i\in V_{1}}{\max}\,x_{i}(T+1)\leq u.
	\label{Eq:4.6.2}
\end{equation}
Repeating (\ref{Eq:4.6.1}) and (\ref{Eq:4.6.2}) continuously, it can be obtained that for any $t\geq T$, $u \geq \max_{i\in V}x_{i}(t)\geq \min_{i \in V}x_{i}(t) $. According to Lemma \ref{L3.5}(i), for $t\geq T$, $ \min_{i\in V}x_{i}(t)$ is monotonically non-decreasing.\\
(ii) The proof is similar to (i).

\subsection{Proof of Lemma \ref{L3.7}}
\noindent(i) Let $\bm{y}(t)=\bm{Ax}(t)$. Since $Q^{*}$ is a weak cohesive group set composed of opinionated agent, according to Definition \ref{D3.3} and (\ref{Eq:4.2.5}), for $\forall\, i\in V$, the following holds
\begin{equation}
	Med_{i}(\bm{y}(t);\bm{M})
	\geq \underset{k\in Q^{*}}{\min}\,y_{k}(t)
	\label{Eq:4.7.-1}.
\end{equation}
Since $\bm{y}(t)=\bm{Ax}(t)$, we have $y_{k}(t)=\bm{a}_{k}\bm{x}(t)$, where $\bm{a}_{k}$ is the $k$-th row of matrix $\bm{A}$.
Since each simplex in $Q^{*}$ is composed of opinionated agent, it follows that $a_{kj}=0$ for $\forall j\in V_{2}\,,\forall\, k\in Q^{*}$.
Also, since the matrix $\bm{A}$ is a row-stochastic matrix, we have $\bm{a}_{k}\bm{1}_{n}=1$. Therefore, it can be deduced that
\begin{equation*}
	y_{k}(t)
	= \bm{a}_{k}\bm{x}(t)
	\geq \bm{a}_{k}(\underset{j\in V_{1}}{\min}\,x_{j}(t)\bm{1}_{n})
	= \underset{j\in V_{1}}{\min}\,x_{j}(t),~\forall\, k \in Q^{*},
\end{equation*} 
then it can be further derived that
\begin{equation*}
	\underset{k \in Q^{*}}{\min}\,y_{k}(t) \geq \underset{j\in V_{1}}{\min}\,x_{j}(t).
\end{equation*}
Therefore, (\ref{Eq:4.7.-1}) can be further derived, for $\forall\, i \in V_{2}\subset V$ satisfying
\begin{equation}
	Med_{i}(\bm{y}(t);\bm{M})
	\geq \underset{j\in V_{1}}{\min}\,x_{j}(t)
	\label{Eq:4.7.1}.
\end{equation}
According to the lemma conditions: $K_G$ does not contain a cohesive agent set consisting only of unopinionated agents, we can obtain that $V_{2}$ and all its subsets are not cohesive agent set. To prove the lemma, the proof proceeds in steps as follows:\\
\\
$step\,1$: $V_{2}$ is not a cohesive agent set, then according to Definition \ref{D3.1}, there exists an agent $h_{1}\in V_{2}$ such that 
\begin{equation*}
	\sum_{j\in V_{2}}w_{h_{1}j}<\frac{1}{2}.
\end{equation*}
Furthermore, we can get 
\begin{equation*}
	\sum_{j\in V_{1}}w_{h_{1}j}=1-\sum_{j\in V_{2}}w_{h_{1}j}>\frac{1}{2}.
\end{equation*}
Since $h_{1}\notin V_{1}$, according to Lemma \ref{L3.2}, we can get
\begin{equation}
	Med_{h_{1}}(\bm{x}(t);\bm{W})
	\geq \underset{j\in V_{1}}{\min}\,x_{j}(t)
	\label{Eq:4.7.2}.
\end{equation}
Then, from (\ref{Eq:4.7.1}) and (\ref{Eq:4.7.2}), we can deduce
\begin{align*}
	&(1-\gamma_{h_{1}})Med_{h_{1}}(\bm{x}(t-1);\bm{W})+\gamma_{h_{1}}Med_{h_{1}}(\bm{y}(t-1);\bm{M})\\
	\geq &(1-\gamma_{h_{1}})\underset{j\in V_{1}}{\min}\,x_{j}(t-1)+\gamma_{h_{1}}\underset{j\in V_{1}}{\min}\,x_{j}(t-1)\\
	\geq &\underset{j\in V_{1}}{\min}\,x_{j}(t-1).
\end{align*}
From (\ref{2.1}), we have
\begin{equation}
	x_{h_{1}}(t)
	\geq\underset{j\in V_{1}}{\min}\,x_{j}(t-1),~\forall\,t \geq T+1.
	\label{Eq:4.7.3}
\end{equation}
$step\,2$: Since $V_{2}\,\backslash \,\{h_{1}\}$ is not a cohesive agent set, according to Definition \ref{D3.1}, there exists an agent $h_{2}\in V_{2}\,\backslash \,\{h_{1}\}$ such that 
\begin{equation*}
	\sum_{j\in V_{2}\backslash \{h_{1}\}}w_{h_{2}j}<\frac{1}{2}.
\end{equation*}
Furthermore, we can get
\begin{equation*}
	\sum_{j\in V_{1}}w_{h_{2}j}+w_{h_{2}h_{1}}=1-\sum_{j \in V_{2}\backslash\{h_{1}\}}w_{h_{2}j}>\frac{1}{2}.
\end{equation*}
Since $h_{2}\notin V_{1}\cup\{h_{1}\}$, according to Lemma \ref{L3.2}, we can get
\begin{equation}
	Med_{h_{2}}(\bm{x}(t);\bm{W})
	\geq\underset{j \in V_{1}}{\min}\,x_{j}(t)\,\land\,x_{h_{1}}(t)    
	\label{Eq:4.7.4}.
\end{equation}
Then, from (\ref{Eq:4.7.1}), (\ref{Eq:4.7.4}), (\ref{Eq:4.7.3}) and the lemma conditions, we can deduce
\begin{align*}
	&(1\!-\!\gamma_{h_{2}})Med_{h_{2}}(\bm{x}(t \!-\!1);\bm{W})\!+\!\gamma_{h_{2}}Med_{h_{2}}(\bm{y}(t \!-\!1);\bm{M})\\
	\geq&(1\!-\!\gamma_{h_{2}})[\underset{j \in V_{1}}{\min}\,x_{j}(t-1)\!\land\! x_{h_{1}}(t \!-\!1)]\!+\!\gamma_{h_{2}}\underset{j\in V_{1}}{\min}\,x_{j}(t \!-\!1)\\
	\geq&\underset{j \in V_{1}}{\min}\,x_{j}(t \!-\!1)\!\land\![(1\!-\!\gamma_{h_{2}})\underset{j\in V_{1}}{\min}\,x_{j}(t \!-\!2)+\gamma_{h_{2}}\underset{j\in V_{1}}{\min}\,x_{j}(t \!-\!1)]\\
	\geq&\underset{j \in V_{1}}{\min}\,x_{j}(t \!-\!1)\!\land\![(1 \!-\!\gamma_{h_{2}})\underset{j\in V_{1}}{\min}\,x_{j}(t \!-\!2)\!+\!\gamma_{h_{2}}\underset{j\in V_{1}}{\min}\,x_{j}(t \!-\!2)]\\
	\geq&\underset{\substack{j\in V_{1}\\t-2\leq s \leq t-1}}{\min}\,x_{j}(s).
\end{align*}
From (\ref{2.1}), we have
\begin{equation*}
	x_{h_{2}}(t)\geq\underset{\substack{j\in V_{1}\\t-2\leq s \leq t-1}}{\min}\,x_{j}(s),~\forall\,t \geq T+2.
\end{equation*}
Repeating the above process.\\
\\
$step\,i$: Since $V_{2}\,\backslash \,\{h_{1},h_{2},...,h_{i-1}\}$ is not a cohesive agent set, then according to Definition \ref{D3.1}, there exists an agent $h_{i}\in V_{2}\,\backslash \,\{h_{1},h_{2},...,h_{i-1}\}$ such that
\begin{equation*}
	\sum_{j\in V_{2}\backslash \{h_{1},...,h_{i-1}\}}w_{h_{i}j}<\frac{1}{2}.
\end{equation*}
Further, we can get
\begin{equation*}
	\sum_{j\in V_{1}\cup \{h_{1},...,h_{i-1}\}}w_{h_{i}j}=1-\sum_{j \in V_{2}\backslash\{h_{1},...,h_{i-1}\}}w_{h_{i}j}>\frac{1}{2}.
\end{equation*}
Since $h_{i}\notin V_{1}\cup \{h_{1},...,h_{i-1}\}$, according to Lemma \ref{L3.2}, we can get
\begin{equation}
	Med_{h_{i}}(\bm{x}(t);\bm{W})
	\geq\underset{j \in V_{1}}{\min}\,x_{j}(t)\!\land \! x_{h_{1}}(t)\!\land\!...
	\!\land\! x_{h_{i-1}}(t)
	\label{Eq:4.7.5}.
\end{equation}
From (\ref{Eq:4.7.5}) and the results obtained in the previous $i-1$ steps, we can further deduce that
\begin{align}
	&Med_{h_{i}}(\bm{x}(t);\bm{W})\notag\\
	\geq&\underset{j \in V_{1}}{\min}\,x_{j}(t)\land x_{h_{1}}(t)\land...\land x_{h_{i-1}}(t)\notag\\
	\geq&\underset{j \in V_{1}}{\min} \, x_{j}(t)\land\underset{j \in V_{1}}{\min} \, x_{j}(t-1)\land...\land\underset{\substack{j \in V_{1}\\t-(i-1)\leq s \leq t-1}}{\min} \, x_{j}(s) \notag\\
	\geq&\underset{j \in V_{1}}{\min} \, x_{j}(t)\land\underset{\substack{j \in V_{1}\\t-(i-1)\leq s \leq t-1}}{\min}\,x_{j}(s).	
	\label{Eq:4.7.5.5}
\end{align}
Then, from (\ref{Eq:4.7.1}) and (\ref{Eq:4.7.5.5}), and the lemma conditions, it can be deduced that
\begin{align*}
	&(1\!-\!\gamma_{h_{i}})Med_{h_{i}}(\bm{x}(t \!-\!1);\bm{W})\!+\!\gamma_{h_{i}}Med_{h_{i}}(\bm{y}(t \!-\!1);\bm{M})\\
	\geq&(1\!-\!\gamma_{h_{i}})[\underset{j \in V_{1}}{\min}\,x_{j}(t \!-\!1)\!\land\!\underset{\substack{j \in V_{1}\\t \!-\! i\leq s \leq t \!-\!2}}{\min}\,x_{j}(s)]\!+\!\gamma_{h_{i}}\underset{j\in V_{1}}{\min}\,x_{j}(t \!-\!1)\\
	\geq&\underset{j \in V_{1}}{\min}\,x_{j}(t \!-\!1)\!\land\![(1\!-\!\gamma_{h_{i}})\underset{\substack{j \in V_{1}\\t \!-\!i\leq s \leq t \!-\!2}}{\min}\,x_{j}(s)\!+\!\gamma_{h_{i}}\underset{\substack{j \in V_{1}\\t \!-\! i\leq s \leq t \!-\!2}}{\min}\,x_{j}(s)]\\
	\geq&\underset{j \in V_{1}}{\min}\,x_{j}(t \!-\!1)\!\land\!\underset{\substack{j \in V_{1}\\t-i\leq s \leq t-2}}{\min}\,x_{j}(s)\\
	\geq&\underset{\substack{j \in V_{1}\\t-i\leq s \leq t-1}}{\min}\,x_{j}(s).
\end{align*}
From (\ref{2.1}), we have
\begin{equation}
	x_{h_{i}}(t)
	\geq \underset{\substack{j\in V_{1}\\t-i\leq s \leq t-1}}{\min}\,x_{j}(s),~\forall\,t \geq T+i
	\label{Eq:4.7.6}.
\end{equation}
Therefore, $h_{1},h_{2},...,h_{n_{2}}$ successively selected from $V_{2}$ all satisfy (\ref{Eq:4.7.6}), which indicates that
\begin{equation*}
	x_{i}(t)
	\geq\underset{\substack{j\in V_{1}\\t-n_{2}\leq s \leq t-1}}{\min}\,x_{j}(s),
	~\forall \, 1\leq i \leq n_{2},\,t \geq T+n_{2}.
\end{equation*}
is proved.\\
(ii) The proof is similar to (i).

\subsection{Proof of Lemma \ref{L3.8}}
\noindent(i) Use mathematical induction to prove that (\ref{Eq:4.8.1}) holds for $\forall\, K\in\mathbb{Z}^{+}$.\\
When $K=1$, according to (\ref{2.1}) and Lemma \ref{L3.1}, we have
\begin{align*}
	&x_{i}(t)-u\\
	\geq&(1-\lambda_{i})[(1-\gamma_{i})\,\underset{j\in V}{\min}\,x_{j}(t-1)+\gamma_{i}\,\underset{j\in V}{\min}\,x_{j}(t-1)-u]\\
	\geq&(1-\lambda_{\max})[\underset{j\in V}{\min}\,x_{j}(t-1)-u]\\
	\geq&(1-\lambda_{\max})[\underset{j\in V}{\min}\,x_{j}(T)-u],~\forall\,i\in V_{1},t\geq T+1.
\end{align*}
Suppose that (\ref{Eq:4.8.1}) holds when $K\leq L$.\\
According to Lemma \ref{L3.7}, we can get
\begin{equation}
	x_{i}(t)
	\geq \underset{\substack{j\in V_{1}\\t-n_{2}\leq s \leq t-1}}{\min}\,x_{j}(s),
	~\forall\,t\geq T+n_{2},\, i\in V_{2}
	\label{Eq:4.8.2}.
\end{equation}
Furthermore, from (\ref{Eq:4.8.2}), we can get
\begin{align*}
	&\underset{j\in V}{\min}\,x_{j}(t)\\
	=&\underset{j\in V_{1}}{\min}\,x_{j}(t)\land\underset{j\in V_{2}}{\min}\,x_{j}(t)\\
	\geq&\underset{j\in V_{1}}{\min}\,x_{j}(t)\land\underset{\substack{j\in V_{1}\\t-n_{2}\leq s \leq t-1}}{\min}\,x_{j}(s)\\
	=&\underset{\substack{j\in V_{1}\\t-n_{2}\leq s \leq t}}{\min}\,x_{j}(s).
\end{align*}
That is
\begin{equation}
	\underset{j\in V}{\min}\,x_{j}(t)
	\geq \underset{\substack{j\in V_{1}\\t-n_{2}\leq s \leq t}}{\min}\,x_{j}(s)
	\label{Eq:4.8.3}.
\end{equation}
Let us denote
\begin{equation}
	x_{j^{*}}(t^{*})=\underset{\substack{j\in V_{1}\\t-n_{2}\leq s \leq t}}{\min}\,x_{j}(s)
	\label{Eq:4.8.4}.
\end{equation}
For $\forall\, t\geq L(n_{2}+1)+T$, since $t^{*}\geq t-n_{2}\geq (L-1)(n_{2}+1)+T+1$, and from the previous assumption, when $K=L$, (\ref{Eq:4.8.1}) holds, i.e.,
\begin{equation}
	x_{j^{*}}(t^{*})-u
	\geq(1-\lambda_{\max})^{L}(\underset{j\in V}{\min}\,x_{j}(T)-u)
	\label{Eq:4.8.5}.
\end{equation}
Therefore, from (\ref{Eq:4.8.3}), (\ref{Eq:4.8.4}) and (\ref{Eq:4.8.5}), we can get
\begin{equation}
	\underset{j\in V}{\min}\,x_{j}(t)
	\geq x_{j^{*}}(t^{*})
	\geq(1-\lambda_{\max})^{L}(\underset{j\in V}{\min}\,x_{j}(T)-u)+u
	\label{Eq:4.8.6}
\end{equation}
for $\forall\,~t\geq L(n_{2}+1)+T$.\\
According to (\ref{2.1}) and (\ref{Eq:4.8.6}), and Lemma \ref{L3.1}, we have
\begin{align*}
	&x_{i}(t)-u\\
	\geq&(1-\lambda_{i})[(1-\gamma_{i})\,\underset{j\in V}{\min}\,x_{j}(t-1)+\gamma_{i}\,\underset{j\in V}{\min}\,x_{j}(t-1)-u]\\
	\geq&(1-\lambda_{\max})[\underset{j\in V}{\min}\,x_{j}(t-1)-u]\\
	\geq&(1-\lambda_{\max})[(1-\lambda_{\max})^{L}(\underset{j\in V}{\min}\,x_{j}(T)-u)]\\
	=&(1-\lambda_{\max})^{L+1}(\underset{j\in V}{\min}\,x_{j}(T)-u),\\
	\forall&\,i\in V_{1}\,,\forall\,t\geq L(n_{2}+1)+T+1.
\end{align*}
Up to this point, it has been proven that when $K=L+1$, (\ref{Eq:4.8.1}) holds.
Therefore, for $\forall\,K\in \mathbb{Z}^{+}$, (\ref{Eq:4.8.1}) holds, which completes the proof.\\
(ii) The proof is similar to (i).

\subsection{Proof of Corollary \ref{C4.1}}
\noindent We consider two cases as follows:\\
Case 1: $n > l$\\
We zero-pad the matrix $\bm{M}$ to construct a square matrix $\bm{M}' = (m'_{ij})_{n \times n}$, where
\begin{equation*}
	m'_{ij}=\begin{cases}
		m_{ij}, &j\leq l;\\
		0, &l<j\leq n.
	\end{cases}
\end{equation*}
Zero pad the vectors $\bm{x}$ and $\bm{y}$ to obtain the vector $\bm{x'}=(x_{1},\ldots,x_{n})$ and $\bm{y'}=(y_{1},\ldots,y_{n})$, where

\vspace{7pt}
\quad\quad$x'_{j}=\begin{cases}
	x_{j}, &j\leq l;\\
	0, &l<j\leq n,
\end{cases}$~~
$y'_{j}=\begin{cases}
	y_{j}, &j\leq l;\\
	0, &l<j\leq n.
\end{cases}$
\vspace{7pt}\\
As directly implied by the padding mechanism, we readily derive that
\begin{equation}
	\|\bm{x'}-\bm{y'}\|_{\infty}=\|\bm{x}-\bm{y}\|_{\infty}
	\label{c4.1}.
\end{equation}
Given that the padded matrix $\bm{M}'$ remains row-stochastic, Lemma \ref{L4.1} immediately yields that
\begin{equation}
	\|Med(\bm{x'};\bm{M'})-Med(\bm{y'};\bm{M'})\|_{\infty}\leq\|\bm{x'}-\bm{y'}\|_{\infty}
	\label{c4.2}.
\end{equation}
We next prove the following result:
\begin{equation}
	Med(\bm{x'};\bm{M'})=Med(\bm{x};\bm{M})
	\label{c4.3}.
\end{equation}
Let $x^{*}=Med_{i}(\bm{x};\bm{M})$. 
By the definition of the weighted median, it holds that
\begin{equation*}
	\underset{j:x_{j}<x^{*}}{\sum}m_{ij}\leq \frac{1}{2}\,,\underset{j:x_{j}>x^{*}}{\sum}m_{ij}\leq \frac{1}{2}.
\end{equation*}
Given that the weights assigned to the padded elements $x'_{l+1},\ldots,x'_{n}$ are zero, and as evident from the padding mechanism, $x'_{k} < x^{*}$ for all $k = l+1,\ldots,n$, it follows that
\begin{equation*}
	\underset{t:x'_{t}<x^{*}}{\sum}m'_{it}=\underset{j:x_{j}<x^{*}}{\sum}m_{ij}+\underset{k=l+1}{\overset{n}{\sum}}m'_{ik}\leq \frac{1}{2}+0=\frac{1}{2},
\end{equation*}
\begin{equation*}
	\underset{t:x'_{t}>x^{*}}{\sum}m_{it}=\underset{j:x_{j}>x^{*}}{\sum}m_{ij}\leq\frac{1}{2}.
\end{equation*}
By Definition \ref{D2.1}, it immediately follows that $	Med_{i}(\bm{x'};\bm{M'})=x^{*}$. The aforementioned procedure is valid for all $1 \leq i \leq n$, thereby establishing the validity of (\ref{c4.3}). Analogously, we readily derive that
\begin{equation}
	Med(\bm{y'};\bm{M'})=Med(\bm{y};\bm{M})
	\label{c4.4}.
\end{equation}
We thus conclude, by virtue of (\ref{c4.1}), (\ref{c4.2}), (\ref{c4.3}), (\ref{c4.4}), and Lemma \ref{L4.1}, that (\ref{c}) holds for non-square matrices $\bm{M}$.\\
Case 2: $n < l$\\
We randomly augment the rows of matrix $\bm{M}$ to construct a square matrix $\bm{M}' = (m'_{ij})_{l \times l}$, where
\begin{equation*}
	m'_{ij}=\begin{cases}
		m_{ij}, & i \leq l; \\
		\frac{1}{l}, & l < i \leq n.
	\end{cases}
\end{equation*}
For simplicity, denote
\begin{equation*}
	\bm{\eta} = Med(\bm{x};\bm{M}) - Med(\bm{y};\bm{M}),
\end{equation*}
\begin{equation*}
	\bm{\eta'}=Med(\bm{x};\bm{M'})-Med(\bm{y};\bm{M'}).
\end{equation*}
Note that $\bm{\eta}$ is an $n$-dimensional vector. As evident from the augmentation process, $\bm{\eta}'$ is an $l$-dimensional vector whose first $n$ components coincide with those of $\bm{\eta}$. It thus follows that
\begin{align}
	&\|Med(\bm{x};\bm{M})-Med(\bm{y};\bm{M})\|_{\infty}\notag\\
	\leq & \| Med(\bm{x};\bm{M'})-Med(\bm{y};\bm{M'}) \|_{\infty}
	\label{c4.5}.
\end{align}
We thus conclude, based on (\ref{c4.5}) and Lemma \ref{L4.1}, that (\ref{c}) holds.

\subsection{Proof of Lemma \ref{L4.2}}
\noindent Let $\bm{x}' = \bm{A}\bm{x}$ and $\bm{y}' = \bm{A}\bm{y}$, where $\bm{x}'$ and $\bm{y}'$ are l-dimensional vectors, and $\bm{M}$ is an $n \times l$ random matrix. Leveraging Corollary \ref{C4.1} and the fact that $\bm{A}$ is an $l \times n$ row-stochastic matrix, it follows that
\begin{align*}
	&\|Med(\bm{Ax};\bm{M})-Med(\bm{Ay};\bm{M})\|_{\infty} \\
	\leq&\|\bm{x'}-\bm{y'}\|_{\infty} \\  
	\leq&\|\bm{A}\|_{\infty}\|\bm{x}-\bm{y}\|_{\infty}\\
	\leq&\|\bm{x}-\bm{y}\|_{\infty}..
\end{align*}
This completes the proof.

\subsection{Proof of Corollary \ref{C4.2}}
\noindent  Leveraging (\ref{4}), (\ref{F(x)}), and (\ref{PQ}), it follows that
\begin{equation*}
	F(\bm{x})=\bm{\Lambda}\bm{u}+\bm{Px}+\bm{QAx}.
\end{equation*}
From the proof of Theorem \ref{T4.1}, the limit point $\bm{x}^*$ is also the unique fixed point of the mapping $F(\bm{x})$, i.e., $F(\bm{x}^*) = \bm{x}^*$. We thus have
\begin{equation*}
	\bm{x}^{*} =\bm{\Lambda}\bm{u}+\bm{Px}^{*}+\bm{QAx}^{*}.
\end{equation*}
Lemma 5.3 below establishes that $\bm{I}_n - \bm{P} - \bm{Q}\bm{A}$ is invertible, which in turn yields Corollary \ref{C4.2}.

\subsection{Proof of Lemma \ref{L4.3}}
\noindent Let $\bm{N}=\bm{QA}$. By the definition of matrix multiplication, it follows that $n_{ij}=\sum_{k}q_{ik}a_{kj}$.
We note that $\bm{Q}$ is a matrix with exactly one non-zero entry per row, and all other entries are zero. Let $q_{i\alpha_i}$ denote the unique non-zero entry in the $i$-th row of $\bm{Q}$. It thus follows that $n_{ij} = q_{i\alpha_i} a_{\alpha_i j}$. Consequently, the $i$-th row of matrix $\bm{N}$ is
\begin{equation*}
	(n_{i1},n_{i2},..., n_{in}) = (q_{i\alpha_{i}}a_{\alpha_{i} 1}, q_{i\alpha_{i}}a_{\alpha_{i} 2}, ..., q_{i\alpha_{i}}a_{\alpha_{i} n}).
\end{equation*}
Given this and the fact that $\bm{A}$ is a row-stochastic matrix, the sum of the $i$-th row of $\bm{N}$ is
\begin{align*}
	\sum_{k=1}^{n}n_{ik}=&q_{i\alpha_{i}}(a_{\alpha_{i} 1}+a_{\alpha_{i} 2}+...+a_{\alpha_{i} n})=q_{i\alpha_{i}}.
\end{align*}
The above equation shows that the sum of the $i$-th row of matrix $\bm{N}$ equals $q_{i\alpha_i}$, the unique non-zero entry in the $i$-th row of $\bm{Q}$.\\
We next consider two cases based on the position of the unique non-zero entry in each row of $\bm{P}$:\\
(1) The non-zero entry in the $i$-th row of $\bm{P}$ lies on the diagonal, i.e., $p_{ii}$. We analyze the diagonal and off-diagonal entries of $\bm{I}_n - \bm{P} - \bm{Q}\bm{A}$ for this case.

(a) Diagonal entry:
\begin{equation*}
	1-p_{ii}-n_{ii}
	=1-p_{ii}-q_{i\alpha_{i}}a_{\alpha_{i} i}.
\end{equation*}

(b) Off-diagonal entry: 
\begin{equation*}
	-n_{ij}=-q_{i\alpha_{i}}a_{\alpha_{i} j}.
\end{equation*}
We next calculate the sum of the absolute values of the off-diagonal entries:
\begin{equation*}
	\sum_{j\neq i}n_{ij}
	=\sum_{j\neq i}q_{i\alpha_{i}}a_{\alpha_{i} j}
	=q_{i\alpha_{i}}\sum_{j\neq i}a_{\alpha_{i} j}
	=q_{i\alpha_{i}}(1-a_{\alpha_{i} i}).
\end{equation*}
From the definitions of matrices $\bm{P}$ and $\bm{Q}$, we can derive the absolute value of the diagonal entry and the sum of the absolute values of the off-diagonal entries for each row, respectively, as follows:
\begin{equation*}
	1-p_{ii}-q_{i\alpha_{i}}a_{\alpha_{i} i}=1-(1-\lambda_{i})(1-\gamma_{i})-(1-\lambda_{i})\gamma_{i}a_{\alpha_{i} i},
\end{equation*}
\begin{equation*}
	q_{i\alpha_{i}}(1-a_{\alpha_{i} i})=(1-\lambda_{i})\gamma_{i}(1-a_{\alpha_{i} i}).
\end{equation*}
If $\bm{I}_{n}-\bm{P}-\bm{QA}$ is a strictly diagonally dominant matrix, it satisfies
\begin{align*}
	&1-(1-\lambda_{i})(1-\gamma_{i})-(1-\lambda_{i})\gamma_{i}a_{\alpha_{i} i}>(1-\lambda_{i})\gamma_{i}(1-a_{\alpha_{i} i})\\
	\Rightarrow&1-(1-\lambda_{i})(1-\gamma_{i})>(1-\lambda_{i})\gamma_{i}\\ 
	\Rightarrow&\lambda_{i}>0.
\end{align*}
Given that $\lambda_i > 0$ holds for all $i$, we conclude that $\bm{I}_n - \bm{P} - \bm{Q}\bm{A}$ is a strictly diagonally dominant matrix.\\
(2) When the non-zero entry in the $i$-th row of $\bm{P}$ is off-diagonal, let $p_{it}$ denote this non-zero entry. We analyze the diagonal and off-diagonal entries of $\bm{I}_n - \bm{P} - \bm{Q}\bm{A}$ for this scenario.

(a) Diagonal entry:
\begin{equation*}
	1-n_{ii}=1-q_{i\alpha_{i}}a_{\alpha_{i} i}.
\end{equation*}

(b) Off-diagonal entry: 

\quad (b.1) Entry in the $t$-th column
\begin{equation*}
	-p_{it}-n_{it}=-p_{it}-q_{i\alpha_{i}}a_{\alpha_{i} t}.
\end{equation*}

\quad (b.2) Entry in the $j$-th column ($j \neq t$ and $j \neq i$)
\begin{equation*}
	-n_{ij}=-q_{i\alpha_{i}}a_{\alpha_{i} j}.
\end{equation*}
We then calculate the sum of the absolute values of the off-diagonal entries:
\begin{equation*}
	p_{it}+\sum_{j,\,j\neq i}n_{ij}=p_{it}+q_{i\alpha_{i}}(1-a_{\alpha_{i} i}).
\end{equation*}
From the definitions of matrices $\bm{P}$ and $\bm{Q}$, we can respectively derive the absolute value of the diagonal entry and the sum of the absolute values of the off-diagonal entries for each row, as follows:
\begin{equation*}
	1-q_{i\alpha_{i}}a_{\alpha_{i} i}=1-(1-\lambda_{i})\gamma_{i}a_{\alpha_{i} i},
\end{equation*}
\begin{equation*}
	p_{it}+q_{i\alpha_{i}}(1-a_{\alpha_{i} i})=(1-\lambda_{i})(1-\gamma_{i})+(1-\lambda_{i})\gamma_{i}(1-a_{\alpha_{i} i}).
\end{equation*}
If $\bm{I}_{n}-\bm{P}-\bm{QA}$ is a strictly diagonally dominant matrix, it satisfies
\begin{align*}
	&1-(1-\lambda_{i})\gamma_{i}a_{\alpha_{i} i}>(1-\lambda_{i})(1-\gamma_{i})+(1-\lambda_{i})\gamma_{i}(1-a_{\alpha_{i} i})\\ 
	\Rightarrow&1-(1-\lambda_{i})\gamma_{i}-(1-\lambda_{i})(1-\gamma_{i})>0\\
	\Rightarrow&\lambda_{i}>0.
\end{align*}
Given that $\lambda_i > 0$ holds for all i, we conclude that $\bm{I}_n - \bm{P} - \bm{Q}\bm{A}$ is strictly diagonally dominant.\\
Since strictly diagonally dominant matrices are invertible, we thus conclude that $\bm{I}_n - \bm{P} - \bm{Q}\bm{A}$ is invertible. This completes the proof.

\subsection{Repeat the experimental results}
\begin{figure}[htbp]
	\centering
	\includegraphics[width = 0.49\textwidth]{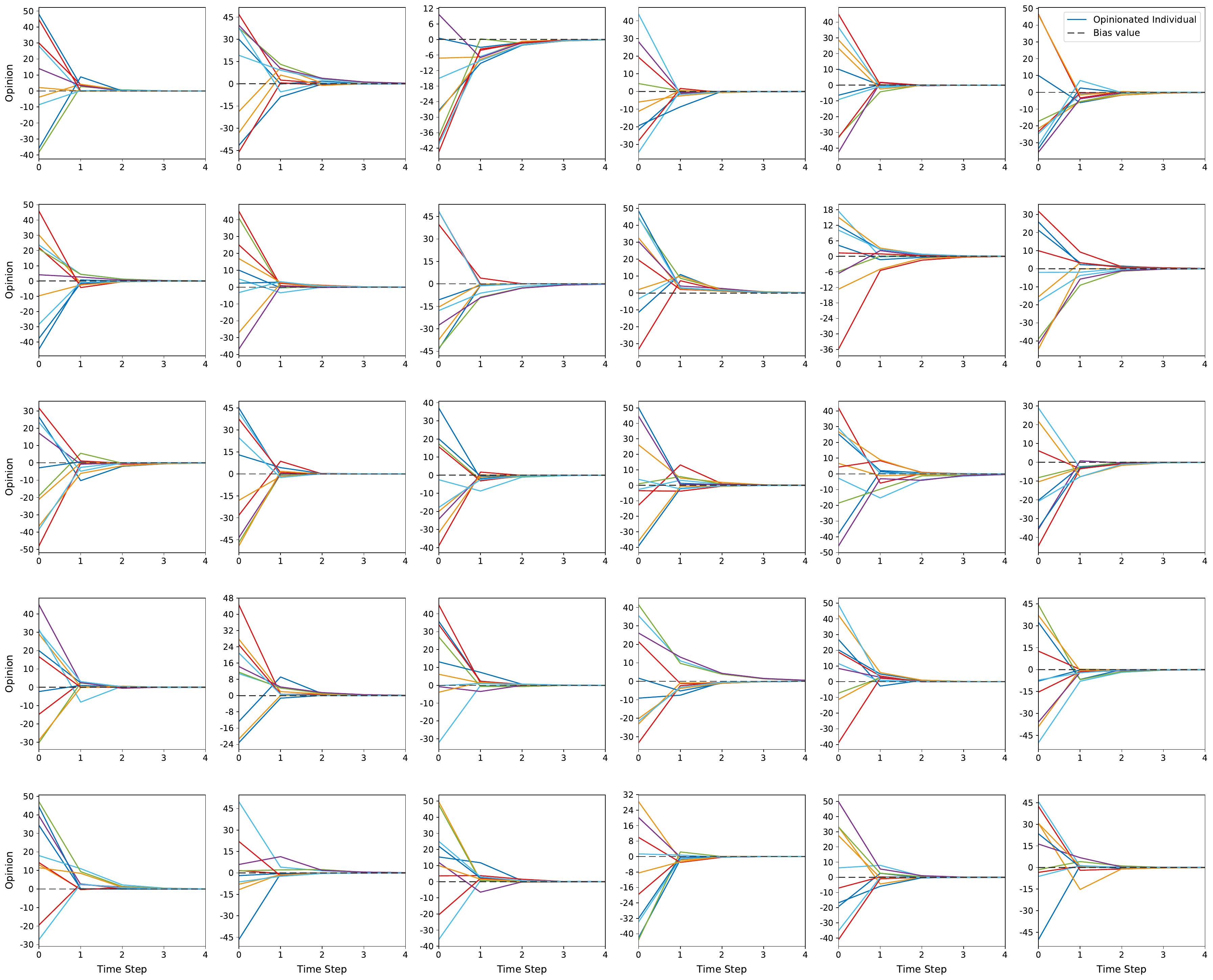}
	\caption{The visualization examples present the changes in agent opinions in repeated experiments with different initial opinion vectors, aiming to eliminate the interference of the coupling between the initial opinion values and the model structure on the experimental results. In the experiments, all agents hold the same bias (the bias value is fixed at 0), and the initial opinion values within the real number range are randomly assigned to 10 agents. The experiments are repeated 30 times (corresponding to 30 subplots in the figure, 5 rows and 6 columns). In each subplot, the horizontal axis represents the time step of opinion evolution, and the vertical axis represents the opinion value of the agent. The solid lines of different colors correspond to the dynamic evolution process of the opinions of the 10 agents, and the gray dashed line represents the same bias value of all agents.}
	\label{rep1}
\end{figure}

\hypertarget{app:L}{}
To exclude potential coupling interference between initial opinion values and the model structure, this work adopts the setting of ``all agents hold the same bias'' to conduct repeated experiments. If it can be verified under this setting that there is no coupling between the selection of initial opinion values and the model structure, then this conclusion can be generalized to any model structure and other agent compositions. In the experiment, we randomly assigned initial opinion values from the real number range to 10 agents and conducted multiple repeated tests. The results show that in the 30 experiments of Fig.\ref{rep1}, all agents eventually reached a consensus, and the consensus was the same bias value of 0, which fully demonstrates that the experimental conclusion is not affected by the specific initial opinion values, further verifying the robustness of the research conclusion.

\bibliographystyle{IEEEtran}
\bibliography{ref}

@article{shen_hybrid_2025,
	title = {A hybrid opinion dynamics model with leaders and followers fusing dynamic social networks in large-scale group decision-making},
	volume = {116},
	issn = {1566-2535},
	doi = {10.1016/j.inffus.2024.102799},
	language = {en},
	journal = {Information Fusion},
	author = {Shen, Yufeng and Ma, Xueling and Deveci, Muhammet and Herrera-Viedma, Enrique and Zhan, Jianming},
	month = apr,
	year = {2025},
	pages = {102799},
}

@article{wang_trust_2024,
	title = {Trust exploration- and leadership incubation- based opinion dynamics model for social network group decision-making: a quantum theory perspective},
	volume = {317},
	issn = {0377-2217},
	doi = {10.1016/j.ejor.2024.03.025},
	language = {en},
	number = {1},
	journal = {European Journal of Operational Research},
	author = {Wang, Peng and Liu, Peide and Li, Yueyuan and Teng, Fei and Pedrycz, Witold},
	month = aug,
	year = {2024},
	pages = {156--170},
}

@article{liu_opinion_2023,
	title = {Opinion dynamics and minimum adjustment-driven consensus model for multi-criteria large-scale group decision making under a novel social trust propagation mechanism},
	volume = {31},
	issn = {1063-6706},
	doi = {10.1109/TFUZZ.2022.3186172},
	language = {en},
	number = {1},
	journal = {IEEE Transactions on Fuzzy Systems},
	author = {Liu, Peide and Li, Yueyuan and Wang, Peng},
	month = jan,
	year = {2023},
	pages = {307--321},
}

@article{zha_opinion_2020,
	title = {Opinion dynamics in finance and business: a literature review and research opportunities},
	volume = {6},
	doi = {10.1186/s40854-020-00211-3},
	language = {en},
	number = {1},
	journal = {Financial Innovation},
	author = {Zha, Quanbo and Kou, Gang and Zhang, Hengjie and Liang, Haiming and Chen, Xia and Li, Cong-Cong and Dong, Yucheng},
	month = dec,
	year = {2020},
}

@article{baumann_modeling_2020,
	title = {Modeling echo chambers and polarization dynamics in social networks},
	volume = {124},
	issn = {0031-9007},
	doi = {10.1103/PhysRevLett.124.048301},
	language = {en},
	number = {4},
	journal = {Physical Review Letters},
	author = {Baumann, Fabian and Lorenz-Spreen, Philipp and Sokolov, Igor M. and Starnini, Michele},
	month = jan,
	year = {2020},
}

@article{dong_survey_2018,
	title = {A survey on the fusion process in opinion dynamics},
	volume = {43},
	issn = {1566-2535},
	doi = {10.1016/j.inffus.2017.11.009},
	language = {en},
	journal = {Information Fusion},
	author = {Dong, Yucheng and Zhan, Min and Kou, Gang and Ding, Zhaogang and Liang, Haiming},
	month = sep,
	year = {2018},
	pages = {57--65},
}

@article{proskurnikov_opinion_2016,
	title = {Opinion dynamics in social networks with hostile camps: consensus vs. {Polarization}},
	volume = {61},
	issn = {0018-9286},
	doi = {10.1109/TAC.2015.2471655},
	language = {en},
	number = {6},
	journal = {IEEE Transactions on Automatic Control},
	author = {Proskurnikov, Anton V. and Matveev, Alexey S. and Cao, Ming},
	month = jun,
	year = {2016},
	pages = {1524--1536},
}

@article{jia_opinion_2015,
	title = {Opinion dynamics and the evolution of social power in influence networks},
	volume = {57},
	issn = {0036-1445},
	doi = {10.1137/130913250},
	language = {en},
	number = {3},
	journal = {SIAM Review},
	author = {Jia, Peng and MirTabatabaei, Anahita and Friedkin, Noah E. and Bullo, Francesco},
	month = jan,
	year = {2015},
	pages = {367--397},
}

@article{french1956formal,
	title = {A formal theory of social power},
	volume = {63},
	language = {en},
	number = {3},
	journal = {Psychological Review},
	author = {French Jr, John RP},
	year = {1956},
	note = {Publisher: American Psychological Association},
	pages = {181},
}

@article{degroot1974reaching,
	title = {Reaching a consensus},
	volume = {69},
	language = {en},
	number = {345},
	journal = {Journal of the American Statistical Association},
	author = {DeGroot, Morris H},
	year = {1974},
	note = {Publisher: Taylor \& Francis},
	pages = {118--121},
}

@article{friedkin1990social,
	title = {Social influence and opinions},
	volume = {15},
	language = {en},
	number = {3-4},
	journal = {Journal of Mathematical Sociology},
	author = {Friedkin, Noah E and Johnsen, Eugene C},
	year = {1990},
	note = {Publisher: Taylor \& Francis},
	pages = {193--206},
}

@article{altafini2012consensus,
	title = {Consensus problems on networks with antagonistic interactions},
	volume = {58},
	language = {en},
	number = {4},
	journal = {IEEE Transactions on Automatic Control},
	author = {Altafini, Claudio},
	year = {2012},
	note = {Publisher: IEEE},
	pages = {935--946},
}

@article{hegselmann2002opinion,
	title = {Opinion dynamics and bounded confidence models, analysis and simulation},
	volume = {5},
	language = {en},
	number = {3},
	journal = {Journal of Artificial Societies and Social Simulation},
	author = {Hegselmann, Rainer and Krause, Ulrich},
	year = {2002},
	note = {Publisher: Journal of Artificial Societies and Social Simulation},
}

@article{deffuant2000mixing,
	title = {Mixing beliefs among interacting agents},
	volume = {3},
	language = {en},
	number = {1n04},
	journal = {Advances in Complex Systems},
	author = {Deffuant, Guillaume and Neau, David and Amblard, Frederic and Weisbuch, Gérard},
	year = {2000},
	note = {Publisher: World Scientific},
	pages = {87--98},
}

@article{dandekar2013biased,
	title = {Biased assimilation, homophily, and the dynamics of polarization},
	volume = {110},
	language = {en},
	number = {15},
	journal = {Proceedings of the National Academy of Sciences},
	author = {Dandekar, Pranav and Goel, Ashish and Lee, David T},
	year = {2013},
	note = {Publisher: National Academy of Sciences},
	pages = {5791--5796},
}

@article{bizyaeva2023nonlinear,
	title = {Nonlinear opinion dynamics with tunable sensitivity},
	volume = {68},
	language = {en},
	number = {3},
	journal = {IEEE Transactions on Automatic Control},
	author = {Bizyaeva, Anastasia and Franci, Alessio and Leonard, Naomi Ehrich},
	year = {2023},
	note = {Publisher: Institute of Electrical and Electronics Engineers Inc.},
	pages = {1415--1430},
}

@article{abelson1964mathematical,
	title = {Mathematical models of the distribution of attitudes under controversy},
	language = {en},
	journal = {Contributions to Mathematical Psychology},
	author = {Abelson, Robert P},
	year = {1964},
	note = {Publisher: Holt, Reinehart and Winston, Inc.},
}

@article{axelrod1997dissemination,
	title = {The dissemination of culture: a model with local convergence and global polarization},
	volume = {41},
	language = {en},
	number = {2},
	journal = {Journal of Conflict Resolution},
	author = {Axelrod, Robert},
	year = {1997},
	note = {Publisher: Sage Periodicals Press 2455 Teller Road, Thousand Oaks, CA 91320},
	pages = {203--226},
}

@article{mei2024convergence,
	title = {Convergence, consensus, and dissensus in the weighted-median opinion dynamics},
	volume = {69},
	language = {en},
	number = {10},
	journal = {IEEE Transactions on Automatic Control},
	author = {Mei, Wenjun and Hendrickx, Julien M and Chen, Ge and Bullo, Francesco and Dörfler, Florian},
	year = {2024},
	note = {Publisher: IEEE},
	pages = {6700--6714},
}

@article{mei2022micro,
	title = {Micro-foundation of opinion dynamics: rich consequences of an inconspicuous change},
	volume = {4},
	language = {en},
	number = {2},
	journal = {Physical Review Research},
	author = {Mei, W and Bullo, F and Chen, G and Hendrickx, J and Dörfler, F},
	year = {2022},
	pages = {23213},
}

@article{zhang2025convergence,
	title = {Convergence analysis of weighted-median opinion dynamics with prejudice},
	language = {en},
	journal = {IEEE Transactions on Automatic Control},
	author = {Zhang, Ruichang and Liu, Zhixin and Chen, Ge and Mei, Wenjun},
	year = {2025},
	note = {Publisher: IEEE},
}

@article{salnikov2018simplicial,
	title = {Simplicial complexes and complex systems},
	volume = {40},
	language = {en},
	number = {1},
	journal = {European Journal of Physics},
	author = {Salnikov, Vsevolod and Cassese, Daniele and Lambiotte, Renaud},
	year = {2018},
	note = {Publisher: IOP Publishing},
	pages = {14001},
}

@article{sizemore2019importance,
	title = {The importance of the whole: topological data analysis for the network neuroscientist},
	volume = {3},
	language = {en},
	number = {3},
	journal = {Network Neuroscience},
	author = {Sizemore, Ann E and Phillips-Cremins, Jennifer E and Ghrist, Robert and Bassett, Danielle S},
	year = {2019},
	note = {Publisher: MIT Press, One Rogers Street, Cambridge, MA 02142-1209, USA},
	pages = {656--673},
}

@article{lambiotte2019networks,
	title = {From networks to optimal higher-order models of complex systems},
	volume = {15},
	language = {en},
	number = {4},
	journal = {Nature Physics},
	author = {Lambiotte, Renaud and Rosvall, Martin and Scholtes, Ingo},
	year = {2019},
	note = {Publisher: Nature Publishing Group UK London},
	pages = {313--320},
}

@article{han2024continuous,
	title = {The continuous-time weighted-median opinion dynamics},
	language = {en},
	journal = {Arxiv Preprint Arxiv:2404.16318},
	author = {Han, Yi and Chen, Ge and Dörfler, Florian and Mei, Wenjun},
	year = {2024},
}

@article{banach1922operations,
	title = {On operations in abstract sets and their application to integral equations},
	volume = {3},
	language = {en},
	number = {1},
	journal = {Fund. Math},
	author = {Banach, Stefan},
	year = {1922},
	pages = {133--181},
}

@article{bassett2012collective,
  title={Collective decision dynamics in the presence of external drivers},
  author={Bassett, Danielle S and Alderson, David L and Carlson, Jean M},
  journal={Physical Review E—Statistical, Nonlinear, and Soft Matter Physics},
  volume={86},
  number={3},
  pages={036105},
  year={2012},
  publisher={APS}
}

@article{melnik2013multi,
  title={Multi-stage complex contagions},
  author={Melnik, Sergey and Ward, Jonathan A and Gleeson, James P and Porter, Mason A},
  journal={Chaos: An Interdisciplinary Journal of Nonlinear Science},
  volume={23},
  number={1},
  year={2013},
  publisher={AIP Publishing}
}

@article{zhuang2019multistage,
  title={Multistage complex contagions in random multiplex networks},
  author={Zhuang, Yong and Ya{\u{g}}an, Osman},
  journal={IEEE Transactions on Control of Network Systems},
  volume={7},
  number={1},
  pages={410--421},
  year={2019},
  publisher={IEEE}
}

@article{carlsson2009topology,
	title = {Topology and data},
	volume = {46},
	language = {en},
	number = {2},
	journal = {Bulletin of the American Mathematical Society},
	author = {Carlsson, Gunnar},
	year = {2009},
	pages = {255--308},
}

@book{aleksandrov1998combinatorial,
	title = {Combinatorial topology},
	volume = {1},
	language = {en},
	publisher = {Courier Corporation},
	author = {Aleksandrov, Pavel S},
	year = {1998},
}

@article{petri2013topological,
	title = {Topological strata of weighted complex networks},
	volume = {8},
	language = {en},
	number = {6},
	journal = {PLOS One},
	author = {Petri, Giovanni and Scolamiero, Martina and Donato, Irene and Vaccarino, Francesco},
	year = {2013},
	note = {Publisher: Public Library of Science San Francisco, USA},
	pages = {e66506},
}

@article{sizemore2017classification,
	title = {Classification of weighted networks through mesoscale homological features},
	volume = {5},
	language = {en},
	number = {2},
	journal = {Journal of Complex Networks},
	author = {Sizemore, Ann and Giusti, Chad and Bassett, Danielle S},
	year = {2017},
	note = {Publisher: Oxford University Press},
	pages = {245--273},
}

@article{kartun2019beyond,
	title = {Beyond the clustering coefficient: a topological analysis of node neighbourhoods in complex networks},
	volume = {1},
	language = {en},
	journal = {Chaos, Solitons \& Fractals:X},
	author = {Kartun-Giles, Alexander P and Bianconi, Ginestra},
	year = {2019},
	note = {Publisher: Elsevier},
	pages = {100004},
}

@article{petri2014homological,
	title = {Homological scaffolds of brain functional networks},
	volume = {11},
	language = {en},
	number = {101},
	journal = {Journal of the Royal Society, Interface},
	author = {Petri, Giovanni and Expert, Paul and Turkheimer, Federico and Carhart-Harris, Robin and Nutt, David and Hellyer, Peter J and Vaccarino, Francesco},
	year = {2014},
	note = {Publisher: The Royal Society},
	pages = {20140873},
}

@article{lord2016insights,
	title = {Insights into brain architectures from the homological scaffolds of functional connectivity networks},
	volume = {10},
	language = {en},
	journal = {Frontiers in Systems Neuroscience},
	author = {Lord, Louis-David and Expert, Paul and Fernandes, Henrique M and Petri, Giovanni and Van Hartevelt, Tim J and Vaccarino, Francesco and Deco, Gustavo and Turkheimer, Federico and Kringelbach, Morten L},
	year = {2016},
	note = {Publisher: Frontiers Media SA},
	pages = {85},
}

@article{lee2012persistent,
	title = {Persistent brain network homology from the perspective of dendrogram},
	volume = {31},
	language = {en},
	number = {12},
	journal = {IEEE Transactions on Medical Imaging},
	author = {Lee, Hyekyoung and Kang, Hyejin and Chung, Moo K and Kim, Bung-Nyun and Lee, Dong Soo},
	year = {2012},
	note = {Publisher: IEEE},
	pages = {2267--2277},
}

@article{sizemore2018cliques,
	title = {Cliques and cavities in the human connectome},
	volume = {44},
	language = {en},
	number = {1},
	journal = {Journal of Computational Neuroscience},
	author = {Sizemore, Ann E and Giusti, Chad and Kahn, Ari and Vettel, Jean M and Betzel, Richard F and Bassett, Danielle S},
	year = {2018},
	note = {Publisher: Springer},
	pages = {115--145},
}

@article{estrada2018centralities,
	title = {Centralities in simplicial complexes. {Applications} to protein interaction networks},
	volume = {438},
	language = {en},
	journal = {Journal of Theoretical Biology},
	author = {Estrada, Ernesto and Ross, Grant J},
	year = {2018},
	note = {Publisher: Elsevier},
	pages = {46--60},
}

@article{sizemore2018knowledge,
	title = {Knowledge gaps in the early growth of semantic feature networks},
	volume = {2},
	language = {en},
	number = {9},
	journal = {Nature Human Behaviour},
	author = {Sizemore, Ann E and Karuza, Elisabeth A and Giusti, Chad and Bassett, Danielle S},
	year = {2018},
	note = {Publisher: Nature Publishing Group UK London},
	pages = {682--692},
}

@article{iacopini2019simplicial,
	title = {Simplicial models of social contagion},
	volume = {10},
	language = {en},
	number = {1},
	journal = {Nature Communications},
	author = {Iacopini, Iacopo and Petri, Giovanni and Barrat, Alain and Latora, Vito},
	year = {2019},
	note = {Publisher: Nature Publishing Group UK London},
	pages = {2485},
}

@article{christy2017theorizing,
  title={Theorizing the impact of targeted narratives: Model admiration and narrative memorability},
  author={Christy, Katheryn R and Jensen, Jakob D and Sarapin, Susan H and Yale, Robert N and Weaver, Jeremy and Pokharel, Manusheela},
  journal={Journal of Health Communication},
  volume={22},
  number={5},
  pages={433--441},
  year={2017},
  publisher={Taylor \& Francis}
}

@article{bal2011influence,
  title={The influence of fictional narrative experience on work outcomes: A conceptual analysis and research model},
  author={Bal, P Matthijs and Butterman, Olivia S and Bakker, Arnold B},
  journal={Review of General Psychology},
  volume={15},
  number={4},
  pages={361--370},
  year={2011},
  publisher={SAGE Publications Sage CA: Los Angeles, CA}
}

@article{dillard2015enhancing,
  title={Enhancing the effects of a narrative message through experiential information processing: An experimental study},
  author={Dillard, Amanda J and Hisler, Garrett},
  journal={Psychology \& Health},
  volume={30},
  number={7},
  pages={803--820},
  year={2015},
  publisher={Taylor \& Francis}
}

@article{kawakami2012implicit,
  title={The implicit influence of a negative mood on the subliminal mere exposure effect},
  author={Kawakami, Naoaki},
  journal={Perceptual and motor skills},
  volume={115},
  number={3},
  pages={715--724},
  year={2012},
  publisher={SAGE Publications Sage CA: Los Angeles, CA}
}

@article{ma2025modeling,
  title={Modeling Opinion Evolution and Conformity Behavior in Large-Scale Social Network Group Decision-Making},
  author={Ma, Xiujuan and Liu, Xinwang and Gong, Zaiwu and Liu, Fang},
  journal={IEEE Transactions on Systems, Man, and Cybernetics: Systems},
  year={2025},
  publisher={IEEE}
}

@article{he2020opinion,
  title={Opinion dynamics with the increasing peer pressure and prejudice on the signed graph},
  author={He, Guang and Zhang, Wenbing and Liu, Jing and Ruan, Haoyue},
  journal={Nonlinear Dynamics},
  volume={99},
  number={4},
  pages={3421--3433},
  year={2020},
  publisher={Springer}
}

@article{semonsen2018opinion,
  title={Opinion dynamics in the presence of increasing agreement pressure},
  author={Semonsen, Justin and Griffin, Christopher and Squicciarini, Anna and Rajtmajer, Sarah},
  journal={IEEE transactions on cybernetics},
  volume={49},
  number={4},
  pages={1270--1278},
  year={2018},
  publisher={IEEE}
}

@inproceedings{amelkin2019fighting,
  title={Fighting opinion control in social networks via link recommendation},
  author={Amelkin, Victor and Singh, Ambuj K},
  booktitle={Proceedings of the 25th ACM SIGKDD international conference on knowledge discovery \& data mining},
  pages={677--685},
  year={2019}
}

@inproceedings{xia2020expressed,
  title={Expressed and private opinion dynamics on influence networks with asynchronous updating},
  author={Xia, Weiguo and Liang, Hong and Ye, Mengbin},
  booktitle={2020 59th IEEE Conference on Decision and Control (CDC)},
  pages={3687--3692},
  year={2020},
  organization={IEEE}
}

\end{document}